\documentclass[review]{elsarticle}

\usepackage{lineno,hyperref}
\usepackage{amsmath,graphicx,booktabs,multirow,textcomp, color}
\usepackage[margin=2.5cm]{geometry}

\modulolinenumbers[5]

\makeatletter
\def\ps@pprintTitle{%
 \let\@oddhead\@empty
 \let\@evenhead\@empty
 \def\@oddfoot{\small{\textit{Postprint of the paper DOI: https://doi.org/10.1016/j.cageo.2018.07.005 published in Computers \& Geosciences}}}%
 \let\@evenfoot\@oddfoot}
\makeatother

\biboptions{authoryear}

\begin{document}

\begin{frontmatter}

\title{PETGEM: A parallel code for 3D CSEM forward modeling using edge finite elements}

\author[mymainaddress]{Octavio Castillo-Reyes}
\cortext[mycorrespondingauthor]{Corresponding author}
\ead{octavio.castillo@bsc.es}

\author[mymainaddress]{Josep de la Puente}
\author[mymainaddress]{Jos\'e Mar\'ia Cela}

\address[mymainaddress]{Barcelona Supercomputing Center (BSC) \\ c/Jordi Girona, 29. 08034 Barcelona, Spain}

\fntext[fn1]{Octavio Castillo-Reyes contributed to the parallel implementation and its documentation.}
\fntext[fn1]{Josep de la Puente contributed to the math background and to the meshes generation.}
\fntext[fn1]{Jos\'e Mar\'ia Cela carried out simulations and its scalability analysis.}

\begin{abstract}
We present the capabilities and results of the Parallel Edge-based Tool for Geophysical Electromagnetic modeling (PETGEM), as well as the physical and numerical foundations upon which it has been developed. PETGEM is an open-source and distributed parallel Python code for fast and highly accurate modeling of 3D marine controlled-source electromagnetic (3D CSEM) problems. We employ the N\'ed\'elec Edge Finite Element Method (EFEM) which offers a good trade-off between accuracy and number of degrees of freedom, while naturally supporting unstructured tetrahedral meshes. We have particularised this new modeling tool to the 3D CSEM problem for infinitesimal point dipoles asumming arbitrarily isotropic media for low-frequencies approximations. In order to avoid source-singularities, PETGEM solves the frequency-domain Maxwell's equations of the secondary electric field, and the primary electric field is calculated analytically for homogeneous background media. We assess the PETGEM accuracy using classical tests with known analytical solutions as well as recent published data of real life geological scenarios. This assessment proves that this new modeling tool reproduces expected accurate solutions in the former tests, and its flexibility on realistic 3D electromagnetic problems. Furthermore, an automatic mesh adaptation strategy for a given frequency and specific source position is presented. We also include a scalability study based on fundamental metrics for high-performance computing (HPC) architectures.
\end{abstract}


\end{frontmatter}

\linenumbers
\nolinenumbers
\section{Introduction}
\label{introduction}
The 3D marine controlled-source electromagnetic method (3D CSEM) is an essential technique in exploration geophysics \citep{Constable2010}. It is based on using an artificial time-varying EM source and analyzing the response at a series of receiver sites. 3D CSEM has been particularly successful finding hydrocarbon reservoirs in offshore exploration, due to hydrocarbon-filled rocks being very resistive (30--500 $\Omega \cdot$m) and surrounded by moderately conductive background media (0.5--2 $\Omega \cdot$m) \citep[see e.g.][]{Weiss2006a, Constable2007, Key2009}. As a key part of the electromagnetic exploration workflow, CSEM surveys for 3D media find two applications. Firstly, modeling allows geophysicists to carry out sensitivity analyses and falsate incorrect models. Secondly, its application is being the engine of an inversion scheme, where 3D CSEM data can be used in order to improve the quality of existing subsurface resistivity models. Although quasi-analytical and 2.5D approaches exist for simplistic scenarios, for really complex geological settings and 3D data surveys, full 3D CSEM modeling tools are necessary.

In particular, 3D CSEM modeling algorithms, like other geophysical modeling approaches, should present the following characteristics:
\begin{enumerate}
\item Efficiency. In terms of being able to obtain accurate results above the noise level expected in the data in a reasonable time at a reasonable cost.
\item Fidelity. Although academic scenarios might be over-simplified, realistic cases often display very complicated geologies. Bathymetry alone can be fairly complex. 
\item Scalability. Schemes must be scalable, easily modifiable and should run equally well on workstations or in compute clusters. 
\end{enumerate}
With these three goals in mind, we present a 3D parallel code for 3D CSEM forward modeling in geophysics, namely, PETGEM: Parallel Edge-based Tool for Geophysical Electromagnetic Modeling. We provide this new tool as open source so that it can be used, modified and redistributed freely with the aims of fostering reproducibility and promoting its use for geophysical 3D CSEM modeling. The code works in parallel for distributed memory computers and it is based upon unstructured tetrahedral meshes for a better representation of geometry due to its simple code structure, it should be fairly easy to upgrade and modify in order to suit the needs of different research applications.

The rest of the paper is organized as follows. In Section 2 we give an overview of 3D CSEM modeling theory and its mathematical formulation. In Section 3 we present the analysis and development of our discretisation approach, namely, N\'ed\'elec Edge Finite Elements (EFEM). In Section 4 we provide a description of the code design and its parallel features. Through comparison with other state of the art algorithms, in Section 5 we describe the application of the code to realistic 3D CSEM models. Finally, we discuss our numerical results and remark PETGEM flexibility, accuracy and capabilities.

\section{3D CSEM forward modeling}
\label{Csem_fm}
In marine 3D CSEM, also referred to as seabed logging \citep{Eidesmo2002}, a deep-towed electric dipole transmitter is used to produce a low-frequency electromagnetic signal which interacts with the electrically conductive Earth and induces eddy currents that become sources of a new electromagnetic signal. The aggregate of both fields is measured by remote receivers placed on the seabed. Since the electromagnetic field at low frequencies, for which displacement currents are negligible, depends mainly on the electric conductivity distribution of the ground, it is possible to detect thin resistive layers beneath the seabed by studying the received signal \citep{Koldan2013}. Operating frequencies of transmitters in 3D CSEM may range between 0.1 and 10 Hz, although in most studies typical frequencies vary from 0.25 to 3 Hz. At such frequencies, and for typical source-receiver offsets of 10--12 km, the penetration depth of the method can extend to several kilometres below the seabed \citep{Hanif2011, Koldan2013}. The main disadvantage of marine 3D CSEM is its relatively low resolution compared to seismic imaging. Therefore, marine 3D CSEM is often used in cooperation with seismic surveying as the latter helps to constrain the geometry of the resistivity model. Marine 3D CSEM is nowadays a well-known geophysical prospecting tool in the industry~\citep[see, e.g., ][]{Constable2006,Boulaenko2007,Constable2007,Orange2009,Constable2010}. 

We consider a quasi-static approximation of the electromagnetic field and ignore the displacement current \citep{Zhdanov2009}. The magnetic permeability of the Earth can be approximated by its value in the free space \citep{Cai2017}. Therefore, the electromagnetic 3D field in an unbounded domain $\Gamma$ can be obtained by solving Maxwell's equations in their diffusive form \citep{Zhdanov2009}
\begin{align}
	\boldsymbol{\nabla} \times \mathbf{E} &= i\omega \mu_{0}\mathbf{H}, \label{eq:maxwell_diffusive_form1} \\
	\boldsymbol{\nabla} \times \mathbf{H} &= \mathbf{J}_{\text{s}} + \tilde{\sigma}\mathbf{E}, \label{eq:maxwell_diffusive_form2}
\end{align}
where we have ommitted the harmonic time dependence e$^{-i \omega t}$. Above, $\omega$ is the angular frequency, $\mathbf E$ the electric field, $\mathbf H$ the magnetizing field, $\mu_{0}$ the free space magnetic permeability, $\mathbf J_{\text{s}}$ the distribution of source current, $\tilde{\sigma}\mathbf E$ the induced current in the conductive Earth and $\tilde{\sigma}$ the electrical conductivity, which is assumed isotropic for simplicity. After substituting eq.~\eqref{eq:maxwell_diffusive_form1} into eq.~\eqref{eq:maxwell_diffusive_form2}, we obtain
\begin{align}
\boldsymbol{\nabla} \times \boldsymbol{\nabla} \times \mathbf{E} + i \omega \mu_{0} \tilde{\sigma} \mathbf{E} = i \omega \mu_{0} \mathbf{J_{\text{s}}},
\label{eq:electric_field_weak_form1}
\end{align}
which is known as the curl-curl form of the problem \citep{Newman2002}. One technique to solving eq.~\eqref{eq:electric_field_weak_form1} is to switch to a primary/secondary field formulation in order to capture the rapid change of the primary field without large grid refinement requirements \citep{Cai2014}. In this case, also referred to as scattering formulation \citep{Zhdanov2009,Cai2014}, the total electric field $\mathbf E$ is obtained as
\begin{align}
\mathbf E &= \mathbf{E_{p}} + \mathbf{E_{s}}, \\
\tilde{\sigma} &=  \tilde{\sigma_{s}} + \Delta \tilde{\sigma},
\label{eq:total_electric_field}
\end{align}
where subscripts $\mathbf{p}$ and $\mathbf{s}$ represent a primary field and secondary field, respectively. For a general layered Earth model, $\mathbf{E_{p}}$ generated upon the simplified conductivity model $\sigma_s$ can be computed semi-analytically by using Hankel transform filters. For homogeneous background media, the development of $\mathbf{E_{p}}$ can be found in~\ref{Appendix_A}. The equation system that must be solved in this case is
\begin{align}
\boldsymbol{\nabla} \times \boldsymbol{\nabla} \times \mathbf{E_{s}} + i \omega \mu_{0} \tilde{\sigma} \mathbf{E_{s}} = -i \omega \mu_{0} \Delta \tilde{\sigma} \mathbf{E_{p}.}
\label{eq:electric_field_weak_form3}
\end{align}
For our modeling purposes we set homogeneous Dirichlet boundary conditions, $\mathbf E = 0$ on $\partial\Gamma$. The range of applicability of this conditions can be determined based on the skin depth of the electric field \citep{Puzyrev2013}.

\section{Edge finite element theory}
\label{Efem_theory}
The most popular discretisation techniques for eq.~\eqref{eq:electric_field_weak_form1} are the Finite Difference Method (FDM) and the Finite Element Method (FEM). FDM is nowadays the most widely employed discretisation scheme \citep[see, e.g., ][]{Alumbaugh1996,Newman2002}. There exist many successful FDM implementations, but the most practical and highly efficient parallel code was developed by \citet{Alumbaugh1996} and improved afterwards by some collaborators. However, the main disadvantage of FDM is its dependance on structured grids, which reduces or limits its accuracy and usability in cases where irregular and complicated geology has a significant influence on measurements. For instance, an imprecise representation of the seabed bathymetry could produce artefacts in images that can lead to false interpretation \citep{Koldan2013}. 

On the other hand, FEM supports completely unstructured meshes as well as mesh refinement, which enables the representation of complex geometries and thus improves the solution accuracy. Nevertheless, FEM is still not as widely applied as FDM and a major obstacle for a wider adoption is that nodal FEM does not correctly take into account all the physical aspects of the vector field functions. In fact, there are three main problems when nodal-based finite elements, obtained by interpolating the nodal values, are employed to represent vector fields (electric or magnetic). The first one is the occurrence of spurious solutions or non-physical solutions, which is generally attributed to the lack of enforcement of the divergence condition \citep{Hiptmair1999, Jin2002, Monk2003, Hiptmair2015}. The second one is the inconvenience of imposing boundary conditions at material interfaces as well as at conducting surfaces \citep{Hiptmair2015}. Finally, the third problem is the difficulty on treating conducting and dielectric edges and corners due to field singularities associated with these structures \citep{Monk2003}. Consequently, most of the researchers who have employed FEM for 3D electromagnetic forward modeling have been primarily focused on overcoming these problems, as well as on solving other physical and numerical challenges, in order to obtain a proper and accurate numerical solution, leaving aside the performance of the codes \citep{Koldan2013}.

Vector basis functions exist that assign vector degrees of freedom (DOF) to the edges rather than to the nodes of each element, which can be used to build so-called Edge Finite elements. In the case of tetrahedral elements and first-order polynomials, a divergence-free basis exists that can be used to build a particular type of edge finite elements called N\'ed\'elec elements. Since N\'ed\'elec elements belong to the class of finite elements method, they are very well suited for honoring complex geometrical structures such as realistic geology or bathymetry. Furthermore, they offer a good trade-off between accuracy and number of DOF, i.e. size of the problem.

\begin{figure}[!htbp]
	\centering
	\includegraphics[width=7cm]{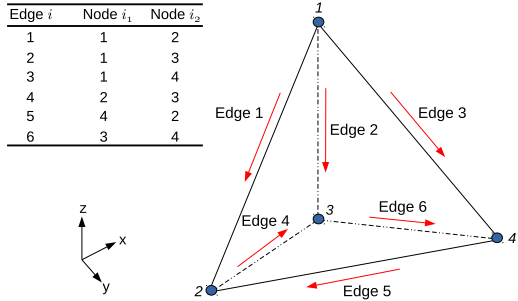}
	\caption{Tetrahedral N\'ed\'elec edge element with node/edge indexing that we have implemented in PETGEM.}
	\label{fig:edge_element}
\end{figure}

For the computation of $\mathbf{E}$ in eq.~\eqref{eq:electric_field_weak_form1}, or alternatively $\mathbf{E_s}$ in eq.~\eqref{eq:electric_field_weak_form3}, we have implemented tetrahedral EFEM of the lowest order which uses vector basis functions defined on the edges of the corresponding elements. These basis functions are divergence-free but not curl-free \citep{Jin2002}. Thus, EFEM naturally ensures tangential continuity and allows normal discontinuity of $\mathbf E$ (or $\mathbf E_{s}$) at material interfaces.  Figure \ref{fig:edge_element} shows the tetrahedral N\'ed\'elec elements together with their node and edge indexing, that we have implemented.
In our approach, we assign the tangential component of the electric field to the mesh edges. Therefore, all components of the electric field at a point $(x, y, z)$ located inside a tetrahedral element $e$ can be obtained as follows
\begin{align}
\mathbf{E}^{e}(x, y, z) = \sum_{i=1}^{6} \mathbf N^{e}_{i}(x,y,z) E^{e}_{i},
\label{eq:field_components_edge_element}
\end{align}
where $\mathbf N^{e}_{i}$ are the vector basis functions associated to each edge $i$, and $E^{e}_{i}$ their respective DOF (see \ref{Appendix_B} for the mathematical development). 

In the following, we show explicitly the development for solving $\mathbf{E_s}$. By substituting eq.~\eqref{eq:field_components_edge_element} into eq.~\eqref{eq:electric_field_weak_form1}, and using Galerkin's approach, the weak form of the original differential equation becomes
\begin{align}
Q_{i} = \int_{\Omega} \mathbf N_{i} \cdot [ \boldsymbol{\nabla} \times \boldsymbol{\nabla} \times \mathbf{E_{s}} -i \omega \mu_{0} \tilde{\sigma} \mathbf{E_{s}} + i \omega \mu_{0} \Delta \tilde{\sigma} \mathbf{E_{p}} ] dV.
\label{eq:electric_field_weak_form2}
\end{align}
The compact discretized form of eq.~\eqref{eq:electric_field_weak_form2} is obtained after applying the Green's theorem
\begin{align}
[K^{e}_{jk} + i \omega \tilde{\sigma}_{e} M^{e}_{jk}] \cdot  \{ E_{sk} \} = - i \omega \mu \Delta \tilde{\sigma}_{e} R^{e}_k,
\label{eq:system_eq_edge_electric}
\end{align}
where $K^{e}$ and $M^{e}$ are the elemental stiffness and mass matrices (see \ref{Appendix_B} for mathematical details). These terms can be calculated analytically or numerically \citep{Jin2002} whereas $R^{e}_k$ is the right hand side that requires numerical integration.

\section{PETGEM: code work-flow}
\label{Petgem_workflow}
\begin{figure}[!htbp]
	\centering
	\includegraphics[width=6.5cm]{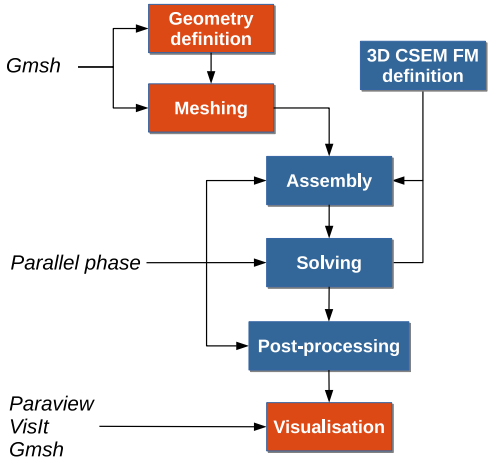}
	\caption{Outline of the overall PETGEM work-flow. On shared-memory architectures, the parallel phase is based on the Multiprocessing package. On the other hand, the \textit{petsc4py} and \textit{mpi4py} packages are used to provide parallel support on distributed-memory architectures.}
	\label{fig:petgem_workflow}
\end{figure}

\begin{figure}[!htbp] 
	\centering    
	\includegraphics[width=11cm]{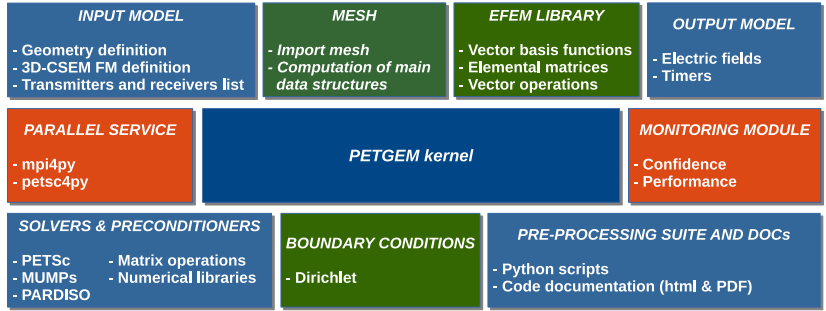}
	\caption{Upper view of PETGEM software stack.}
	\label{fig:petgem_sw_stack}
\end{figure}

PETGEM is a Python 3 code for the scalable solution of 3D CSEM on tetrahedral meshes, as these are the easiest to scale-up to very large meshes of arbitrary shape. It is written mostly in Python 3 and relies on the scientific Python software stack with use of \textit{mpi4py} ~\citep{Dalcin2011} and \textit{petsc4py} ~\citep{Dalcin2011} packages for parallel computations. Other scientific Python packages used include: \textit{Numpy} for efficient array manipulation and \textit{Scipy} algorithms for numerical computations. PETGEM allow users the simulation of electromagnetic fields in realistic 3D CSEM on distributed-memory HPC platforms. Among others, the key drivers for the PETGEM development are the following:
\begin{enumerate}
\item Fill the relative scarcity of robust edge-based codes for 3D CSEM to reduce ambiguities in data interpretation for hydrocarbon exploration.
\item Model realistic scenarios that support 3D CSEM simulations in structurally complex geometries with a good trade-off between accuracy and number of DOF.
\item Provide synthetic results which can then compared to measured data.
\item Improve the degree integration of HPC using Python, EFEM, and geophysical methods such as 3D CSEM at realistic-scale.
\end{enumerate}
Commonly the 3D CSEM is composed of four main tasks: discretisation of the geometry, elemental matrices computation, and global system assembly, solving the resulting system and post-processing the solution. The problem decomposition into independent modules is important because each region make use of methods that belong to different branches of mathematics. For instance, the design of algorithms for 3D meshing and iterative solvers for large scale modeling require knowledge going beyond the scope of this work. Therefore, this paper rely on well-known tools for domain discretisation and solving systems of linear equations and focuses on the kernel of 3D CSEM, namely, the core of PETGEM. An outline of the overall PETGEM work-flow is depicted in Figure \ref{fig:petgem_workflow}. Furthermore, in Figure \ref{fig:petgem_sw_stack} a snapshot of the primary groups of modules in the code is given. 
In order to solve a 3D CSEM case, PETGEM requires a parameters file with all information associated with the model under consideration. For the sake of simplicity and in order to avoid a specific parser, the PETGEM parameters file is defined as a Python dictionary. This parameters file is divided into four sections: physical parameters (frequency, source position, source current, source length, conductivity model, background conductivity), mesh information (file path of nodal spatial coordinates, nodal element connectivity, edge element connectivity, edges nodes connectivity, and sparsity structure for \textit{PETSc} matrix allocation ~\citep{petsc2016}), solver parameters (solver and preconditioner type, tolerance, maximum number of iterations), results information (receivers position file path). Regarding discretisation formats, PETGEM is capable of importing tetrahedral finite element mesh files generated by \textit{Gmsh}~\citep{Geuzaine2008}. 
Notice that geometry and meshing modules are independent of main PETGEM work-flow.

Once the parameter file is defined, PETGEM imports it and starts the assembly of the global linear system. Several processes of the kernel can be spawned, each responsible for its own subdomain, so that the whole domain is covered. Each process then assembles its local contributions to the global linear system that is solved. For this purpose, PETGEM use the \textit{PETSc} library and its large collection of data structures and parallel iterative solvers, that can be used in Python through the \textit{petsc4py} and \textit{mpi4py} packages. 

Once a solution of the 3D CSEM has been obtained, it should be post-processed by using a visualization program. PETGEM does not the visualization by itself, but it generates output files (\textit{PETSc}, \textit{MATLAB}, and \textit{ASCII} format) with the electric field responses that can be easily imported by external visualization tools.

PETGEM is written in Python 3 because it is open source and functional on a wide number of platforms, including HPC environments. Furthermore, it uses a high level and very expressive language. The code structure is modular, simple, and flexible which allows exploiting not just PETGEM modules but also third party libraries. The HPC goal of this code involves using cutting-edge architectures. To that goal, the code is implemented in current state-of-the-art platforms such as Intel Skylake, Intel Haswell and Intel Xeon Phi processors, which offer high performance, flexibility, and power efficiency. Nevertheless, PETGEM support older architectures such as SandyBridge, for the sake of usability and to be able to compare performance.

\section{Application scenarios}
In order to verify our EFEM formulation and study the PETGEM capabilities, we simulate 3D CSEM forward modeling over different scenarios. These experiments have been performed on version III of the \textit{Marenostrum} supercomputer at  \textit{BSC}. Marenostrum (MN3) supercomputer based on Intel SandyBridge processors, iDataPlex Compute Racks, a Linux Operating System and an Infiniband interconnection. It has $48\:896$ Intel SandyBridge-EP E5–2670 cores at 2.6 GHz grouped into $3\:056$ computing nodes, 103.5 TB of main memory (128 nodes with 128 Gb, 128 nodes 64 Gb, and $2\:880$ nodes with 32 Gb) as well as 1.9 PB of GPFS disk storage. Its peak perfomance is 1.1 Petaflops. Each computing node has two 8-core Intel Xeon processors E5-2670 with a frequency of 2.6 GHz and 20 MB cache memory.

Regarding numerical accuracy, we defined a relative misfit criteria for the electric responses in terms of its amplitude $|\mathbf{E}|$ and phase $\Phi$ as follows
\begin{align}
\epsilon &= \left|\frac{(S^{ref}-S^{efem})}{S^{ref}}\right| \times 100,
\label{eq:misfits}
\end{align}
where $S^{ref}$ and $S^{efem}$ are the reference and EFEM solutions, respectively. 

\subsection{Canonical model of an off-shore hydrocarbon reservoir}
\label{DIPOLE1D_test}
\begin{figure}[!htbp]
	\centering
	\includegraphics[width=8.5cm]{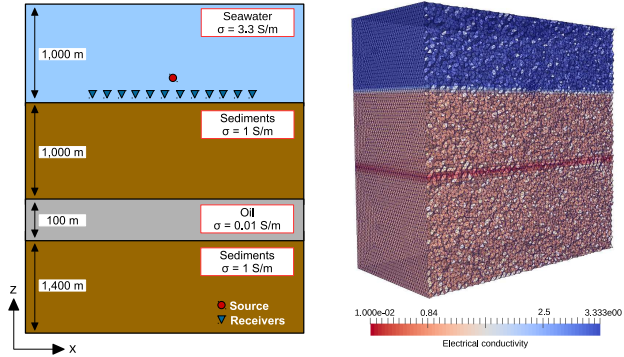}
	\caption{In-line canonical off-shore hydrocarbon model with its unstructured tetrahedral mesh for $y>1\:750$ m. The color scale represents the electrical conductivity $\sigma$ for each layer.}
	\label{fig:model1}
\end{figure}

In the first test, we validate our approach and PETGEM  against the quasi-analytical results of the canonical model by \citet{Weiss2006a}. Our computational domain is defined by a $[-1,4.5]\times[0,3.5]\times[0,3.5]$ km composed by four layers with different thickness and conductivity ($\sigma$): $1\:000$ m thick seawater (3.3 S/m), $1\:000$ m thick sediments (1 S/m), 100 m thick oil (0.01 S/m), and $1\:400$ m thick sediments (1 S/m). Figure \ref{fig:model1} shows a 3D view of the model with its unstructured tetrahedral mesh for the halfspace $y>1\:750$ m, with the color scale representing the electrical conductivity $\sigma$ for each layer.
For this model, we use a 1 Hz $x$-directed dipole source as in \citet{Castillo2015b}, which is located at $x=1\:750$ m, $y=1\:750$ m, and $z=975$ m. The receivers are placed in-line to the source position and along its orientation, directly above the seafloor ($z = 990$) with spacing of $58$ m. For this case, we computed the electric fields using a tetrahedral mesh with $\approx2$ millions of DOF. Also, a \textit{PETSc} implementation of the GMRES solver has been used to solve the system of equations.

\begin{figure}[!htbp]
	\centering
	\includegraphics[width=9cm]{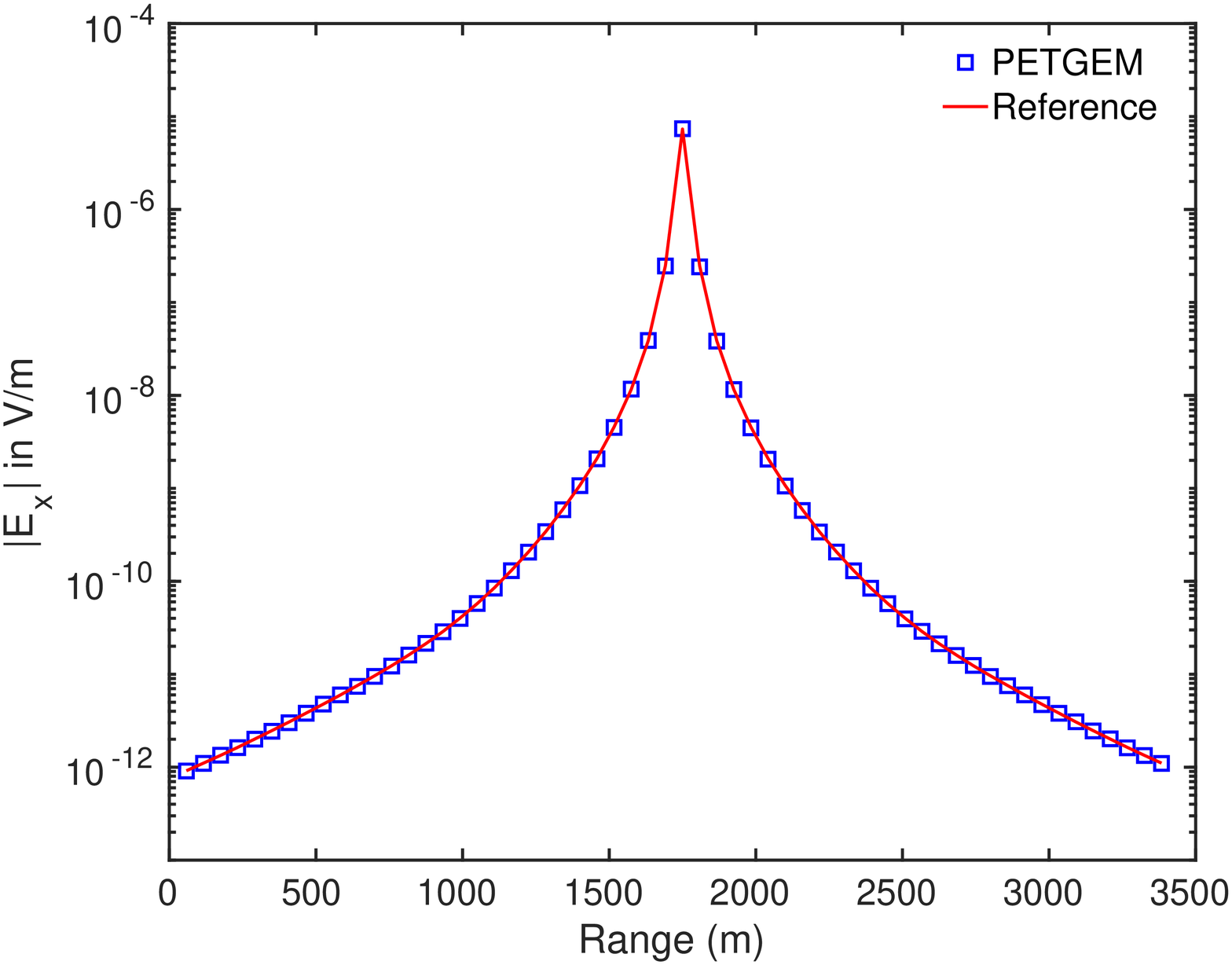}
	\includegraphics[width=9cm]{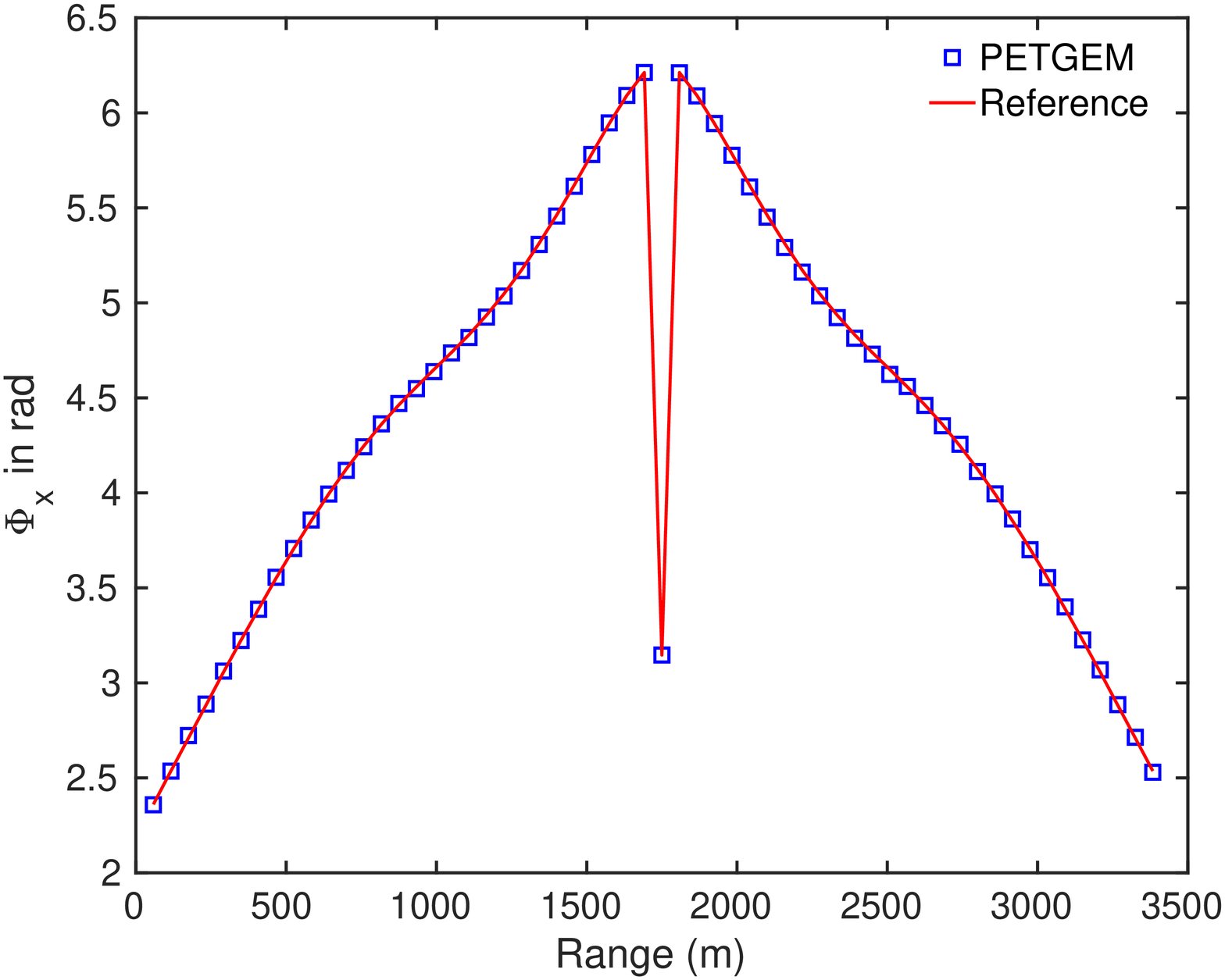}
	\caption{Comparison of inline electric field responses obtained from PETGEM and DIPOLE1D \citep{Key2009}. Amplitude $|\mathbf{E}_{x}|$ and phase $\Phi_{x}$ are plotted at top and bottom, respectively.}
	\label{fig:fields_X_model1}
\end{figure}

\begin{figure}[!htbp]
	\centering
	\includegraphics[width=9cm]{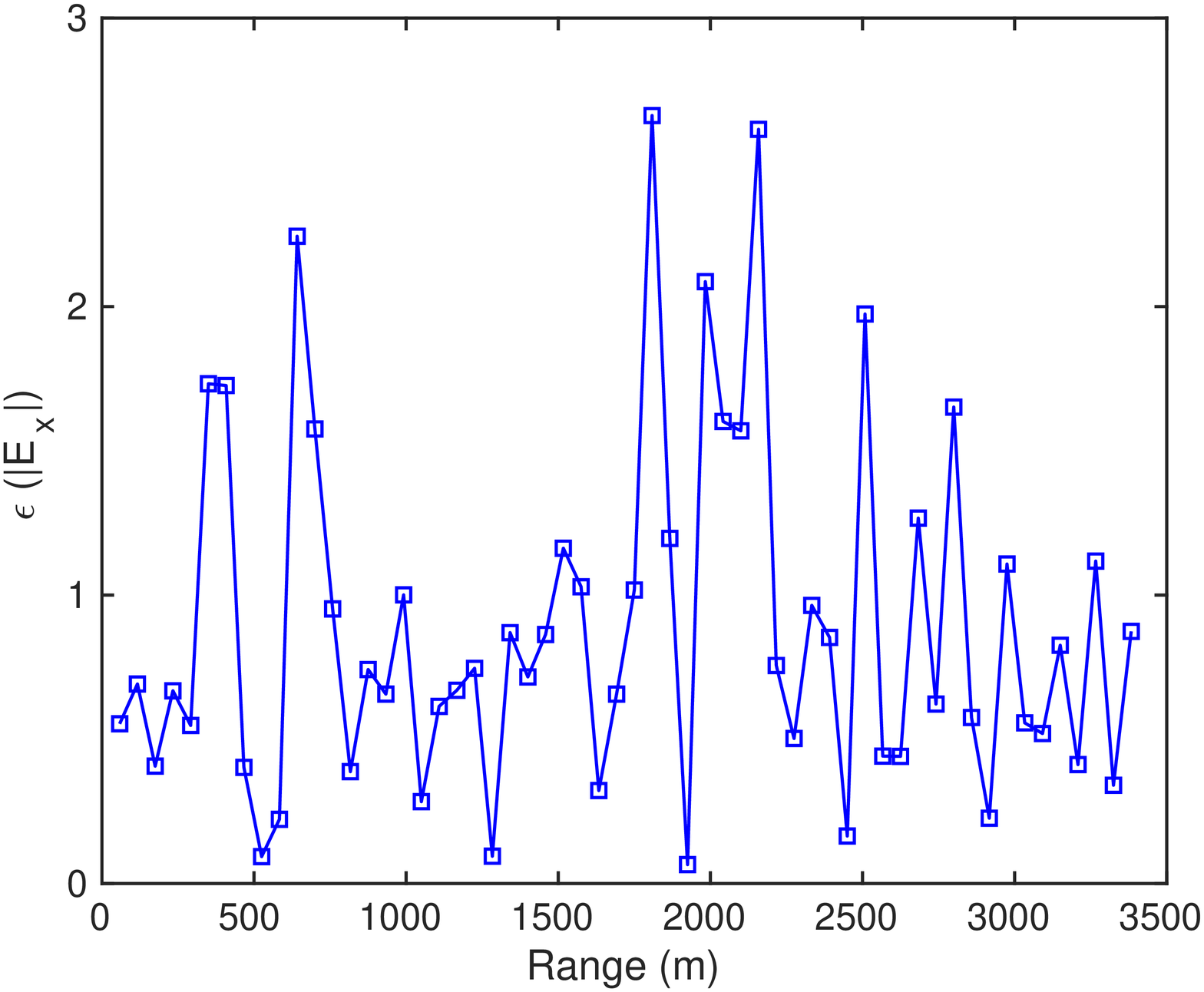}
	\includegraphics[width=9cm]{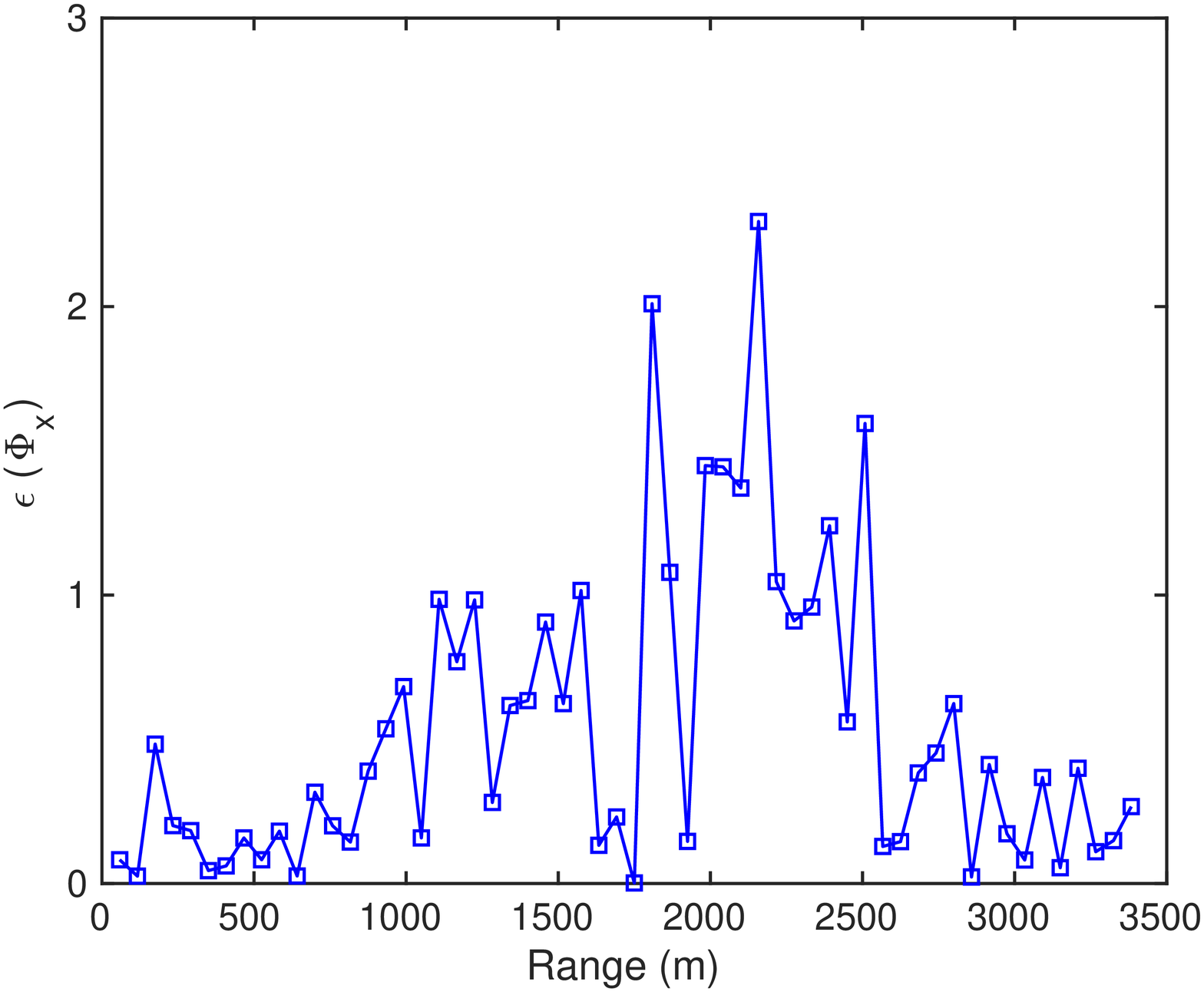}
	\caption{Amplitude misfit $\epsilon(|\mathbf{E}_{x}|)$ and phase misfit $\epsilon(\Phi_{x})$ of $\mathbf{E}_{x}$  shown in Figure \ref{fig:fields_X_model1}. The $\epsilon(|\mathbf{E}_{x}|)$ and $\epsilon(\Phi_{x})$ misfits are plotted at top and bottom, respectively. In both cases, an average relative misfit $<1\%$ is observed.}
	\label{fig:diff_X_model1}
\end{figure}

Figure \ref{fig:fields_X_model1} shows the amplitude and phase comparison of $\mathbf{E}_{x}$ measurements between PETGEM and those obtained with the DIPOLE1D tool \citep{Key2009}. The top panel depicts the amplitude 
$|\mathbf{E}_{x}|$ of the electric responses along the receivers line, where is easy to see that PETGEM results are in good agreement with the reference data. Additionally, the phase ratio $\Phi_{x}$ comparison is presented in the bottom panel of Figure \ref{fig:fields_X_model1}. 
Based on this verification, we computed the amplitude $\epsilon(|\mathbf{E}_{x}|)$ and phase $\epsilon(\Phi_{x})$ misfits for inline receivers. The results are shown in Figure \ref{fig:diff_X_model1}. In both cases, a good overall agreement is observed ($<1\%$ of average relative misfit).
Hence, the PETGEM solution shows a good agreement with the quasi-analytical results in canonical models.

\subsection{CSEM modeling with bathymetry}
\label{Bathymetry_test}
The second test involves a 3D CSEM modeling with bathymetry. This model is especially interesting because a primary advantage of the FEM/EFEM over other techniques like the Finite Diference Method (FDM) is the precise representations of arbitrarily complex geological structures such as seafloor bathymetry, without critically increasing problem size. Furthermore, if not taken into account, bathymetry effects can produce large anomalies on the measured electric fields.

\begin{figure}[!htbp]
\centering
\includegraphics[width=5.5cm]{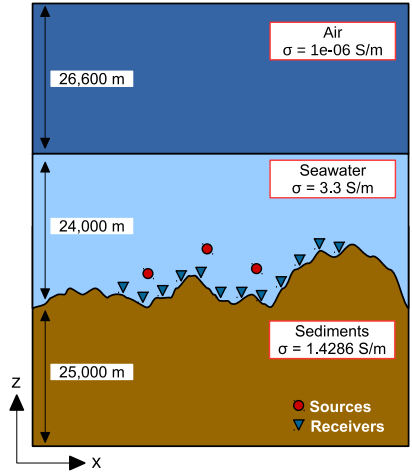}
\caption{Bathymetry model used for simulating 3D CSEM surveys. The model was discretized into $1,542,514$ tetrahedral elements, resulting in $1,814,928$ unknowns. The dataset of this 3D CSEM model was provided by \citet{Chung2014}.}
\label{fig:model2}
\end{figure}

\begin{figure}[!htbp]
\centering
\includegraphics[width=9cm]{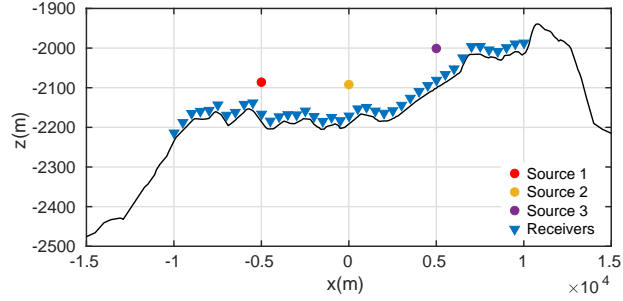}
\caption{Cross-sectional view of the bathymetry model in Figure \ref{fig:model2}. Three $0.25$ Hz $x-$directed dipoles are located at $[-5, 0, -2.086]$ km, $[0, 0, -2.096]$ km, and $[5, 0, -2.001]$ km. The 41 receivers are placed in-line to the source position and alongs its orientation, directly above the seafloor.}
\label{fig:bathymetry}
\end{figure}

\begin{figure}[!htbp]
	\centering
	\includegraphics[width=12cm]{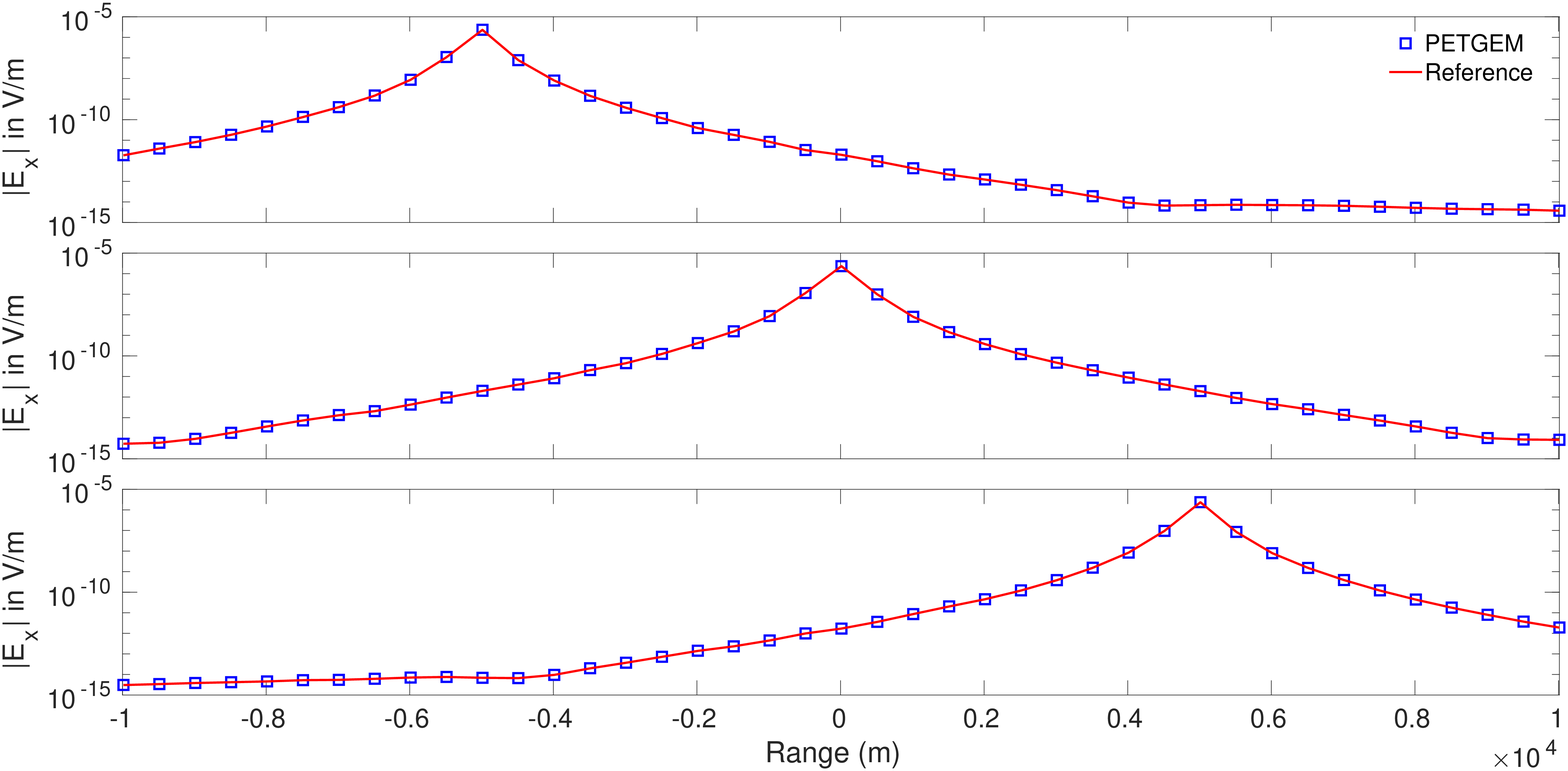}
	\caption{Amplitude $|\mathbf{E}_{x}|$ comparison between PETGEM and the results of \citet{Chung2014}. The three $0.25$ Hz sources located at $[-5, 0, -2.086]$ km, $[0, 0, -2.096]$ km, and $[5, 0, -2.001]$ km are plotted on the top, middle and bottom, respectively.}
	\label{fig:total_field_X_model2}
\end{figure}

\begin{figure}[!htbp]
	\centering
	\includegraphics[width=12cm]{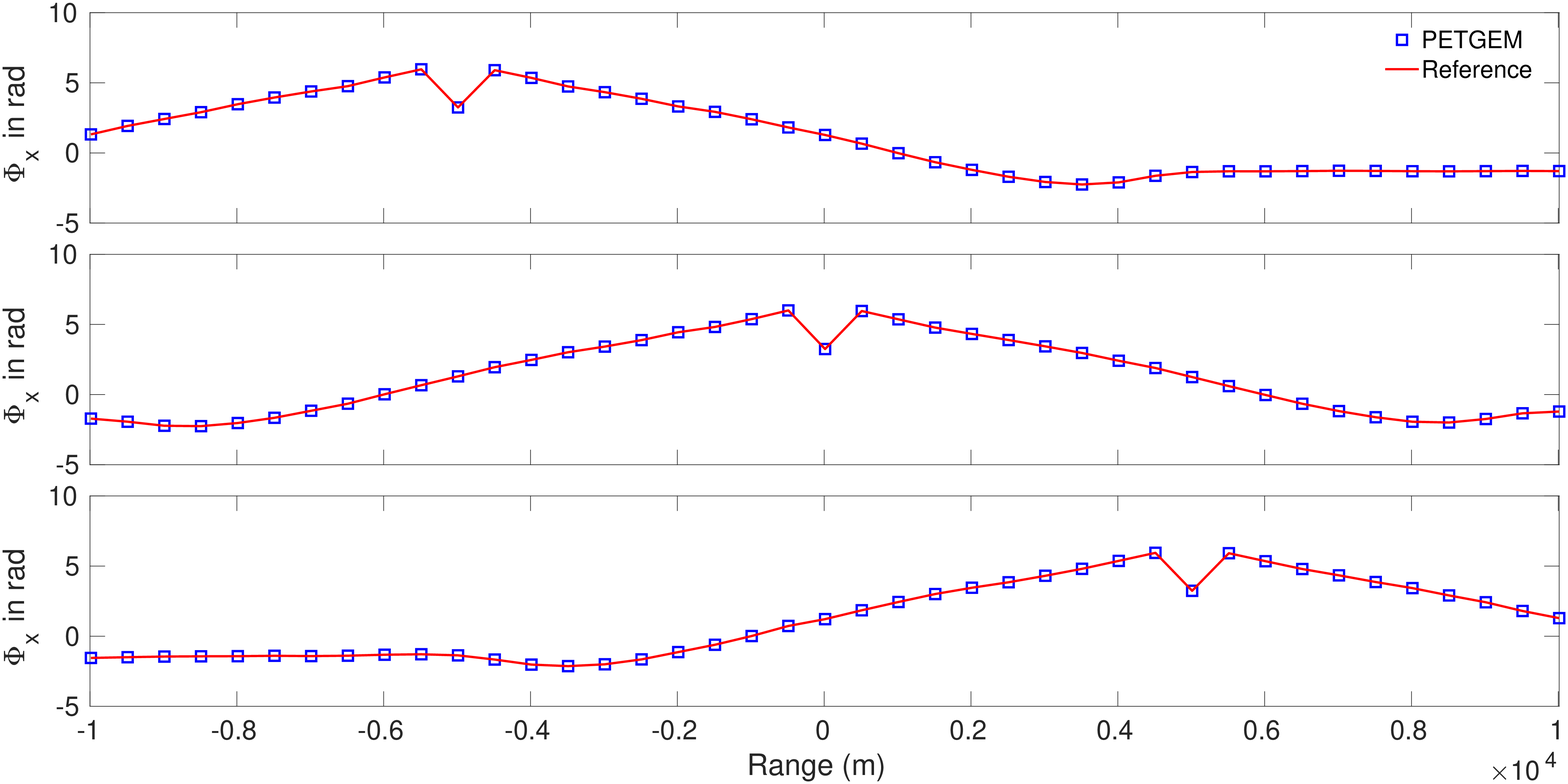}
	\caption{Phase $\Phi_{x}$ comparison between PETGEM and the results of \citet{Chung2014}. The three $0.25$ Hz sources located at $[-5, 0, -2.086]$ km, $[0, 0, -2.096]$ km and $[5, 0, -2.001]$ km are plotted on the top, middle, and bottom, respectively.}
	\label{fig:phase_field_X_model2}
\end{figure}

The reference dataset of this model was provided by \citet{Chung2014}. Additionally, a nodal FEM solution of this modeling case is described in \citet{Um2013}. The model consists of $26.6$ km of air layer (1e-6 S/m), $2.4$ km of seawater (3.3 S/m), and $25.0$ km of sediments layer (1.4286 S/m). The computational domain is defined by a $[0,30]\times[0,32]\times[0,54]$ km box as shown in Figure \ref{fig:model2}. The model has been discretized into $1\:542\:514$ tetrahedral elements, resulting in $258\:716$ nodes and $1\:814\:928$ DOFs.
For this simulation, we use three x-oriented electric dipole sources, with a moment of 200 A$\cdot$m and frequency of $0.25$ Hz, located at $[-5, 0, -2.086]$ km, $[0, 0, -2.096]$ km, and $[5, 0, -2.001]$ km. The 41 receivers are placed in-line to the source positions and along its orientation, directly above the seafloor, as shown in Figure ~\ref{fig:bathymetry}. The bathymetry in the area is very rough, and adequate spatial sampling can only be achieved by allowing receiver deployment in slopes or trenches. Several steep trenches go through the area from the shallow eastern part to the deeper western part. For this model, we use the multifrontal parallel solver MUMPS \citep{MUMPS2006} to solve the resulting system of equations.
For each source, Figures \ref{fig:total_field_X_model2} and \ref{fig:phase_field_X_model2} compare the amplitude $|\mathbf{E}_{x}|$ and phase $\Phi_{x}$ of electric fields obtained from our modeling tool with those produced by \citet{Chung2014}. One can clearly see a good agreement with the reference.

\begin{figure}[!htbp]
	\centering
	\includegraphics[width=12cm]{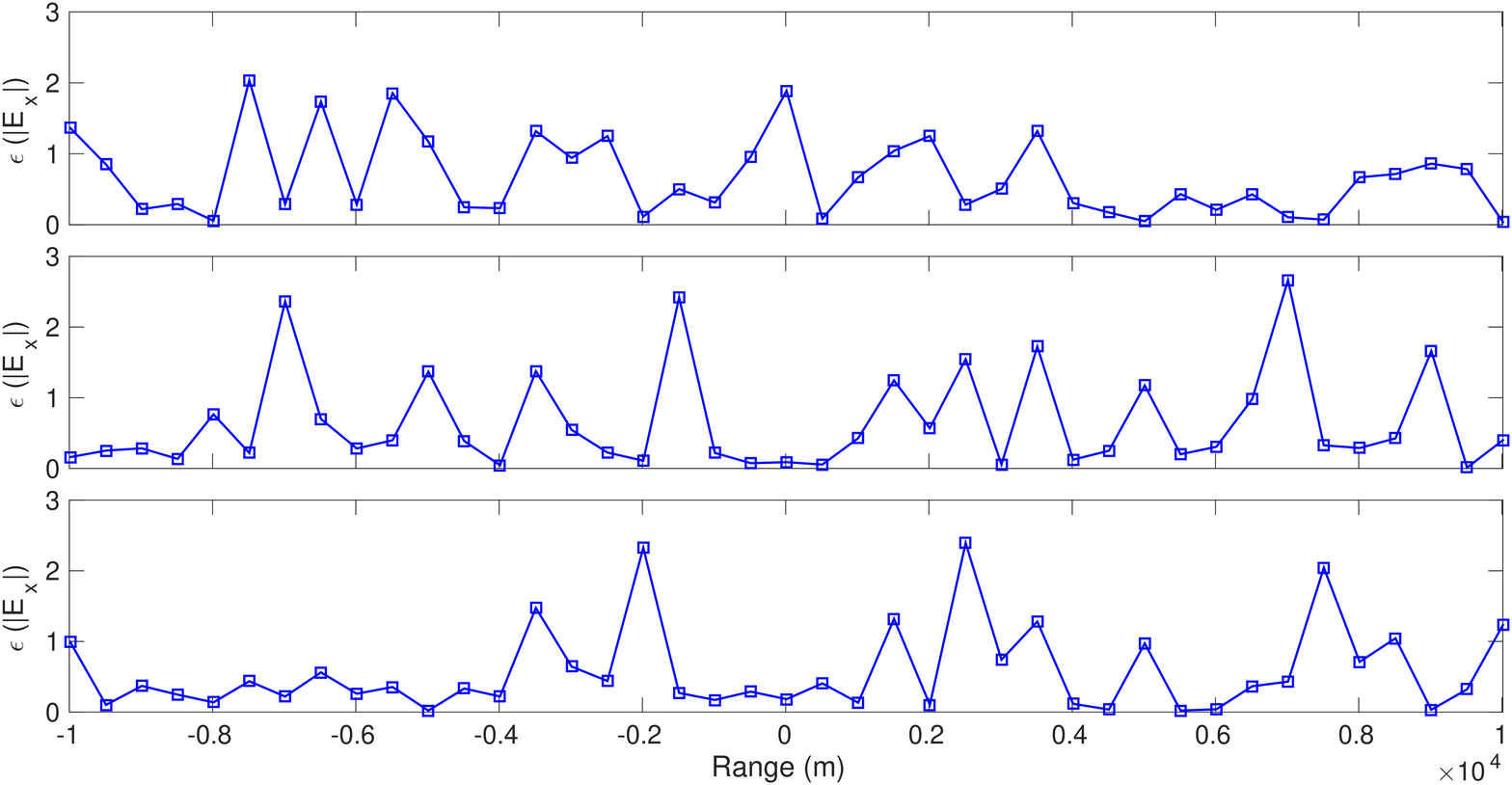}
	\caption{Amplitude misfits $\epsilon(|\mathbf{E}_{x}|)$ of Figure \ref{fig:total_field_X_model2}. The $\epsilon(|\mathbf{E}_{x}|)$ for the three $0.25$ Hz sources located at $[-5, 0, -2.086]$ km, $[0, 0, -2.096]$ km, and $[5, 0, -2.001]$ km are plotted on the top, middle, and bottom, respectively.}
	\label{fig:diff_amplitude_X_model2}
\end{figure}

\begin{figure}[!htbp]
	\centering
	\includegraphics[width=12cm]{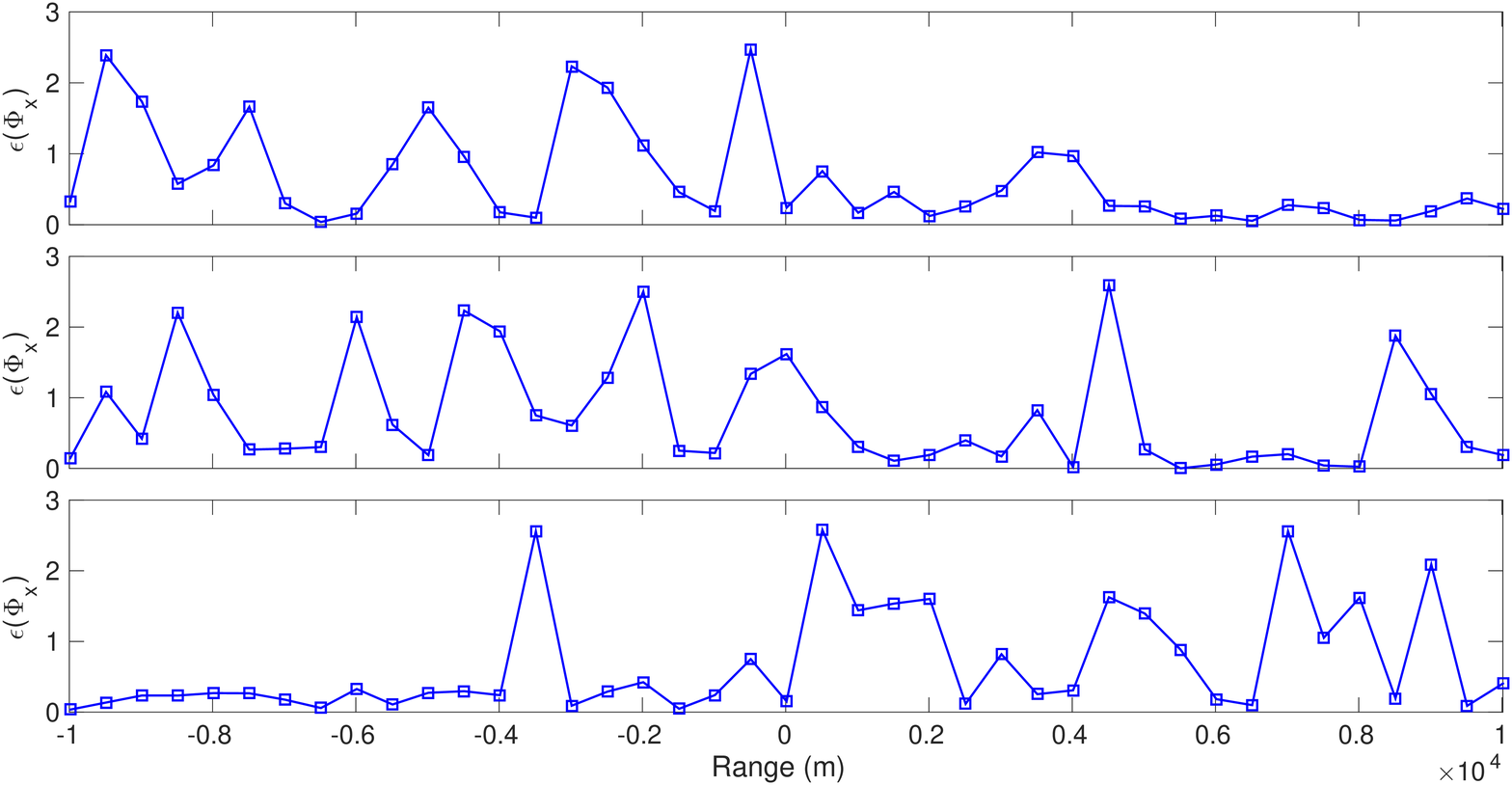}
	\caption{Phase misfits $\epsilon(\Phi_{x})$ of Figure \ref{fig:phase_field_X_model2}. The $\epsilon(\Phi_{x})$ for the three $0.25$ Hz sources located at $[-5, 0, -2.086]$ km, $[0, 0, -2.096]$ km, and $[5, 0, -2.001]$ km are plotted on the top, middle, and bottom, respectively.}
	\label{fig:diff_phase_X_model2}
\end{figure}

As in the previous model, we compare PETGEM solution against the reference in terms of misfit ratios. Figures \ref{fig:diff_amplitude_X_model2} and \ref{fig:diff_phase_X_model2} present the amplitude $\epsilon(|\mathbf{E}_{x}|)$ and phase $\epsilon(\Phi_{x})$ misfits, respectively. The overall agreement is reasonable ($<2\%$ of average relative misfits) when considering that both results are subject to different numerical inaccuracy comming from discretisation method, mesh quality, among others.
On the other hand, in \citet{Chung2014} the solution for the model was computed using EFEM over a hexahedral mesh with $265\times64\times73$ cells and $3\:796\:596$ DOFs. Furthermore, the authors reported that the computation spent $3\:872$ seconds and required less than $88.5$ Gb of memory on one node equipped with two Intel quad-core Xeon processors (resulting in eight cores) at 2.53 GHz and sharing 96 Gb of memory. In order to compare these numbers we executed our simulation with 8 MPI tasks. The mean runtime for this model is $2\:439$ seconds and required less than 64 Gb. This means a good efficiency of PETGEM  when considering the difference between the DOF of each model. Again, PETGEM results show a good overall agreement with the reference solution.

\subsection{Automatic mesh adaptation}
\label{Automatic_mesh_test}
Nowadays, in the field of numerical simulations based on FEM and EFEM, automatic mesh adaptation has largely proved its efficiency for improving the accuracy of the numerical solution and capturing the behavior of physical phenomena by exploiting local mesh refinement. In principle, this technique allowing substantially reducing the number of DOF, thus favorably impacting CPU times, and to achieving a desired accuracy on computed solutions. Although iterative solvers are rather efficient even on oversampled meshes, memory requirements can be reduced if the computational mesh is adapted to the source location and to the frequency. However, when the source and receivers depth and the bathymetry are varying, it will be too cumbersome and impractical to ask the user to define a mesh per source/receivers and per frequency. 

This test is devoted to the analysis of the automatic mesh adaptation approach developed by \citet{Plessix2007}. This approach ensures, for a given frequency and a given source position, that the computational domain is consistent with the discretization of the EM equations. Its core is based on the skin-depth ($\delta$), defined as the effective depth of penetration of EM energy in a conducting medium, where the amplitude of a plane wave in a whole space has been attenuated to $1/e$ or $37\%$ \citep{Sheriff2002}. Its formal definition is the following
\begin{align}
\delta = \sqrt{\frac{2}{\mu_{0}\omega\sigma}} \approx 503 \sqrt{\frac{1}{f \sigma}},
\label{eq:skin_depth}
\end{align}
where $\mu_{0}$ is the free space magnetic permeability (H/m), $\omega$ is the angular frequency (rad), $\sigma$ is the electric conductivity (S/m), and $f$ is the frequency (Hz). According to the formulation of~\citet{Plessix2007}, equation ~\eqref{eq:skin_depth} gives a rule to automatically determine the spacing $d_{\delta}(f)$ at a frequency $f$
\begin{align}
d_{\delta}(f) = \frac{\delta_{\min}(f)}{r_{\delta}},
\label{eq:skin_depth_frequency}
\end{align}
where $\delta_{\text{min}}$ is the minimum skin-depth and $r_{\delta}$ is a number between two and three. In 3D CSEM surveys, $\delta_{\text{min}}$ occurs in the water layer where $\sigma \approx 3.3$ S/m and $\delta_{\text{min}} \approx 275 / \sqrt{f}$. In order to obtain better approximations around source and receivers, we define the spacing $d_{\text{s}}$ as follows
\begin{align}
d_{\text{s}} = \min\left(\frac{L_{\text{s}}}{r_{\text{s}}}, d_{\delta} (f)\right),
\label{eq:source_refinement}
\end{align}
where $L_{\text{s}}$ is the source dipole length and $r_{\text{s}}$ a number between ten and fifteen. The value for $r_{\text{s}}$ is different to those described by \citet{Plessix2007} (between two and four) because the authors used a finite-integration approach. However, in the test described below, we observe that this difference does not imply a significant increase in the computational cost.

We estimate the mesh dimensions from the average skin-depth ($\delta_{ave}$), that generally corresponds to the skin-depth in the sediment areas, i.e., $\delta_{ave} \approx 503 / \sqrt{f}$ for a conductivity of 0.01 S/m. The computational domain, decomposed into a core domain and extra boundary layers, is defined as follows
\begin{align}
[x_{\text{s}}-r_{x}\delta_{ave}, x_{\text{s}} + r_{x}\delta_{ave}] \times [y_{\text{s}}-r_{y}\delta_{ave}, y_{\text{s}} + r_{y}\delta_{ave}] \times [z_{air}, z_{\text{s}} + r_{z}\delta_{ave}],
\label{eq:core_computational_domain}
\end{align}
where $r_{x}$ and $r_{y}$ are numbers between four and eight (depending on the location of the receivers), $r_{z}$ a number around four, $z_{air}$ the depth of the air-water interface, and $[x_{\text{s}},y_{\text{s}},z_{\text{s}}]$ the source position. In order to reduce boundary reflections (as can be seen in Section \ref{DIPOLE1D_test}), we add extra boundary layers with a thickness of $t_{b} = r_{b}\delta_{ave}$, where $r_{b}$ is a number around four. The values of $r_{x}$, $r_{y}$, 
$r_{z}$, and $r_{b}$ where chosen in the same way as by~\citet{Plessix2007}, where reflections of EM fields are reduced by $98\%$ over four skin-depths, and by $99.9\%$ over eight skin-depths.

For this test, the core of the computational domain is centered at $x_{\text{s}}$ and the number of points is limited by the following power-law stretching
\begin{align}
x_{i} = x_{i-1} + \min ( s_{c}^{i} d_{\text{s}}, d_{\delta}(f) ),
\label{eq:power_law_stretching_core}
\end{align}
where $s_{c}$ is the stretching parameter equal to 1.04. Similarly, the boundary layers were stretched with a power-law defined as
\begin{align}
x_{i} = x_{i-1} + s_{b}(x_{i-1} - x_{i-2}),
\label{eq:power_law_stretching_boundaries}
\end{align}
where $s_{b}$ is equal to 1.1. Finally, we have defined a constant conductivity value for each element of the computational domain.

\begin{figure}[!htbp]
	\centering
	\includegraphics[trim=0 55 0 0, clip, width=12cm]{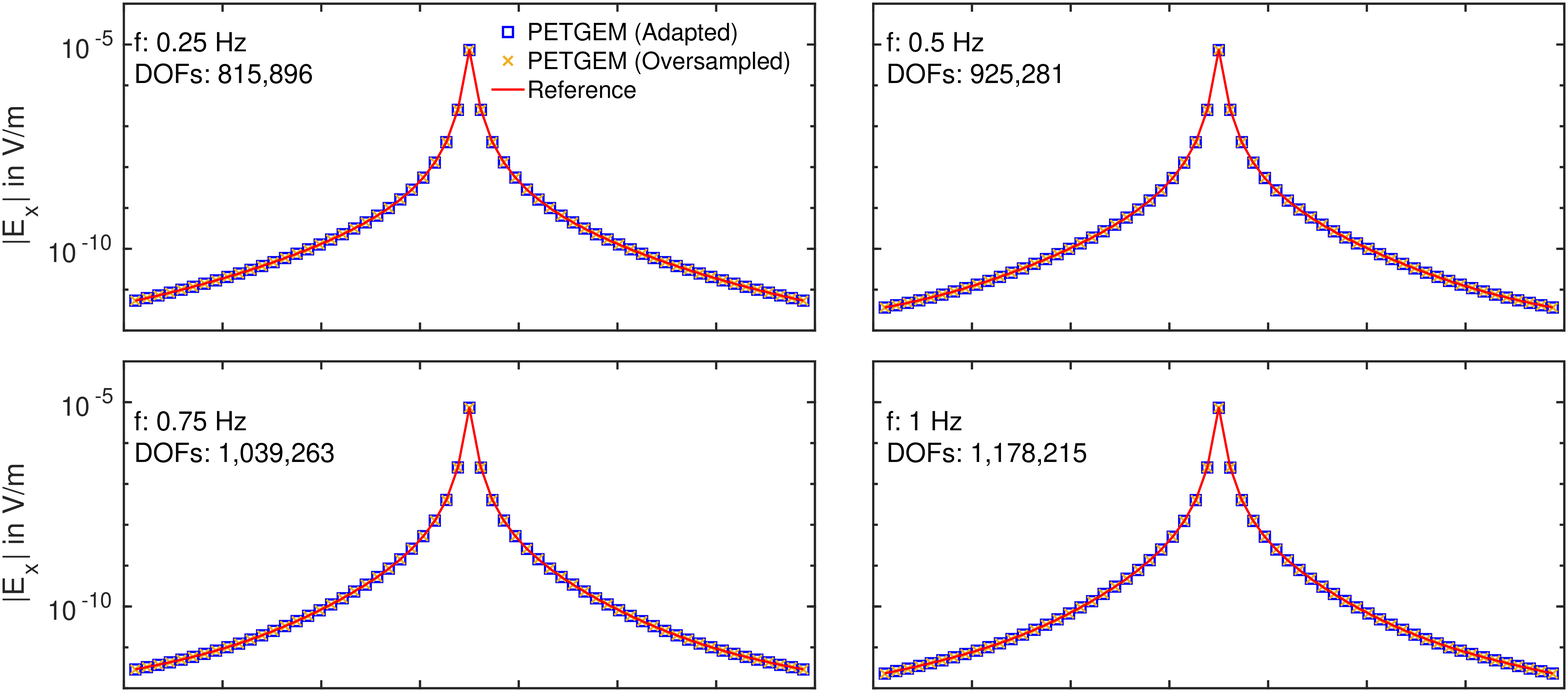}
	\includegraphics[width=12cm]{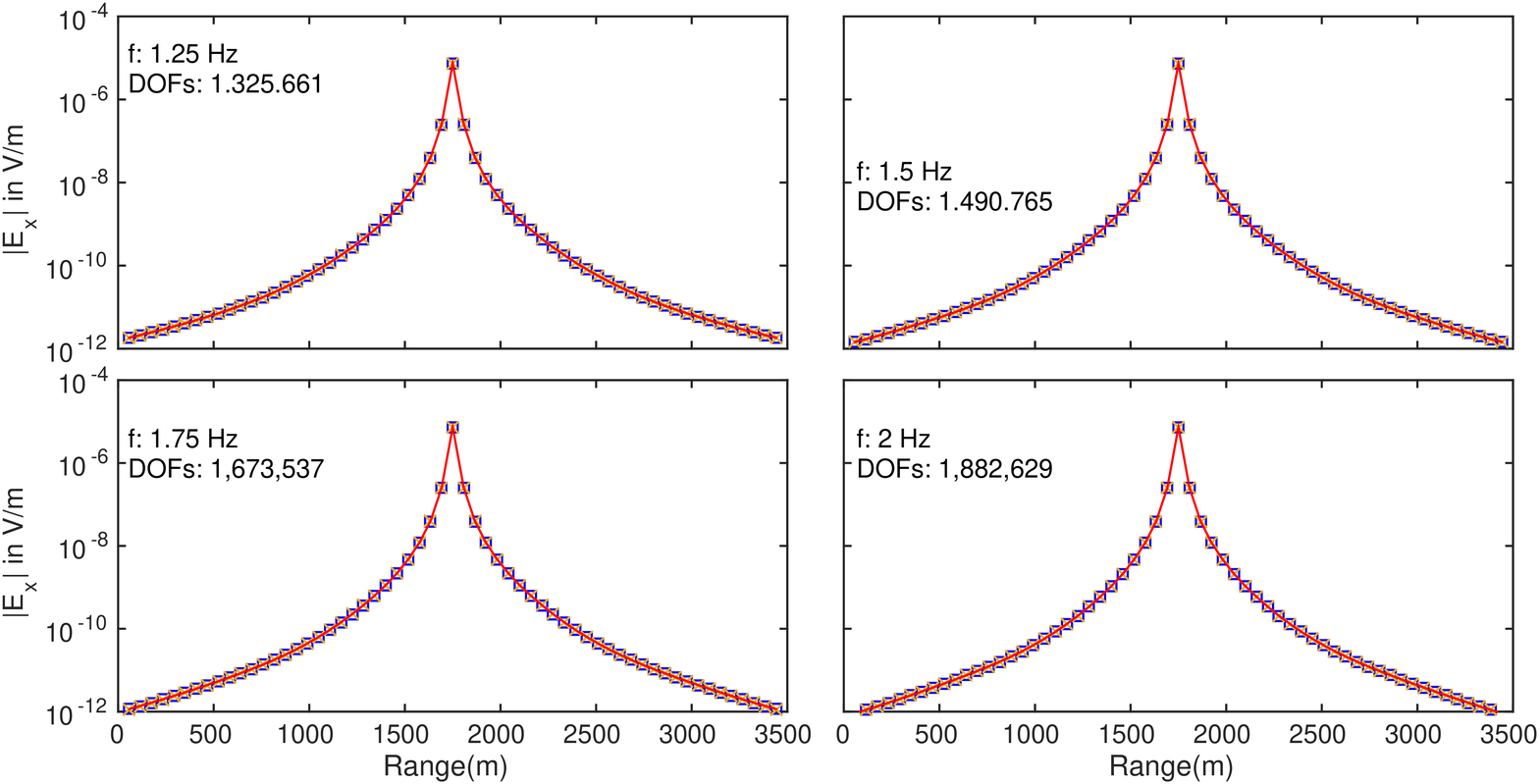}
	\caption{Comparison of inline electric field responses between PETGEM and those obtained with DIPOLE1D \citep{Key2009}. Here, the amplitude $|\mathbf{E}_{x}|$ is plotted for each frequency.}
	\label{fig:total_field_X_model3}
\end{figure}

\begin{figure}[!htbp]
	\centering
	\includegraphics[trim=0 55 0 0, clip, width=12cm]{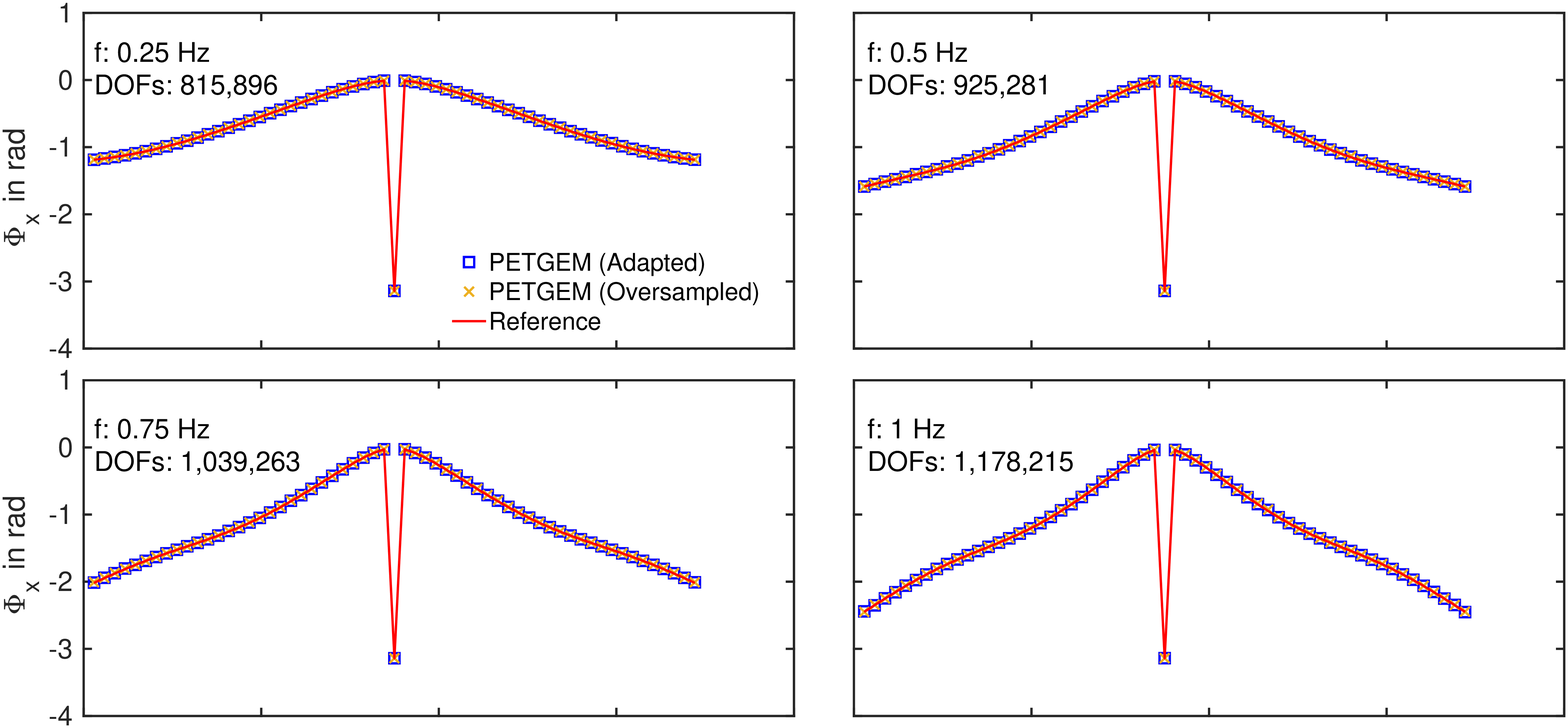}
	\includegraphics[width=12cm]{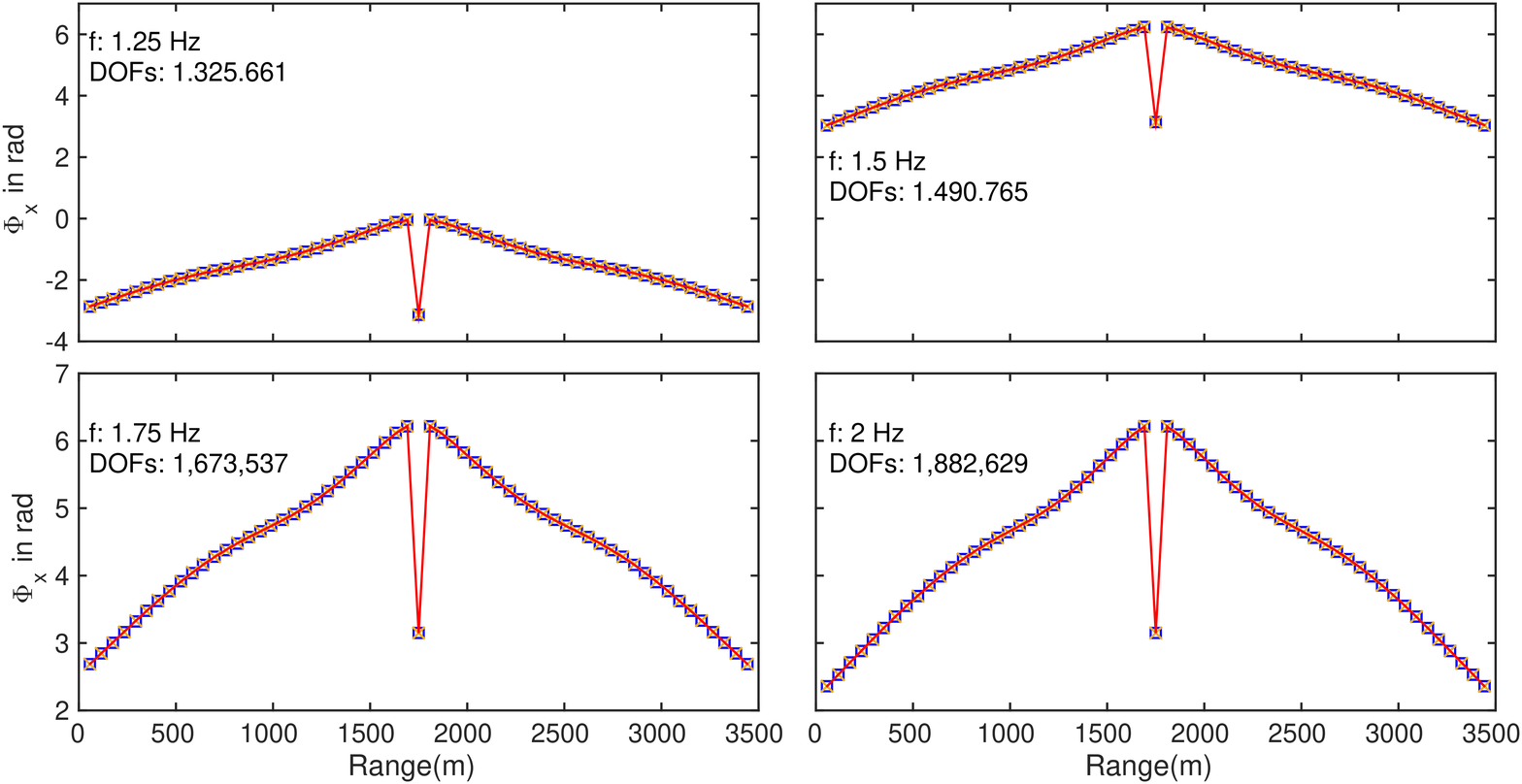}
	\caption{Comparison of inline electric field responses between PETGEM and those obtained with DIPOLE1D \citep{Key2009}. Here, the phase $\Phi_{x}$ is plotted for each frequency.}
	\label{fig:phase_field_X_model3}
\end{figure}

To evaluate whether the aforementioned approach is satisfactory to model 3D CSEM surveys, we carried out several PETGEM  simulations based on the model described in Section \ref{DIPOLE1D_test} for the following frequencies: 0.25 Hz,  0.5 Hz, 0.75 Hz, 1 Hz, 1.25 Hz, 1.5 Hz, 1.75 Hz, and 2 Hz. The strategy was as follows: The geometry of the domain, including the interfaces and separation between core and extra domains is generated beforehand in \textit{Gmsh}~\citep{Geuzaine2008}. Then we set up an automatic meshing script that, given an input frequency, applies the discretization rules aformentioned to define typical spacings to each region. By running the script, \textit{Gmsh} generates a new mesh in a completely automated way. For each frequency, the computational mesh was controlled by the paramters $r_{\delta} = 3$, $r_{\text{s}} = 13$, $r_{x} = r_{y} = 8$, $r_{z} = 4$, and $r_{b} = 4$. A summary of the resulting meshes from this process is described in Table~\ref{table:meshes_automatic_adaptation}. All tests have been solved with a single process using the Symmetric Quasi-Minimal Residual (SQMR) solver with a Successive Over-relaxation (SOR) method as preconditioner.
\begin{table}[!htbp]
\centering
\caption{Summary of resulting meshes based on automatic mesh adaptation.}
\small
\begin{tabular}{cccccc}
\toprule
Label   & Frequency (Hz) & Elements & Nodes & Edges & DOF \\
\midrule
A & 0.25 & $717\:572$ & $122\:461$ & $852\:100$ & $815\:896$ \\ 
B & 0.5  & $811\:710$ & $137\:708$ & $961\:485$ & $925\:281$ \\ 
C & 0.75 & $909\:821$ & $153\:579$ & $1\:075\:467$ & $1\:039\:263$ \\ 
D & 1    & $1\:029\:484$ & $172\:868$ & $1\:214\:419$ & $1\:178\:215$ \\ 
E & 1.25 & $1\:156\:378$ & $193\:420$ & $1\:361\:865$ & $1\:325\:661$ \\ 
F & 1.5  & $1\:298\:556$ & $216\:346$ & $1\:526\:969$ & $1\:490\:765$ \\ 
G & 1.75 & $1\:455\:924$ & $241\:750$ & $1\:709\:741$ & $1\:673\:537$ \\ 
H & 2    & $1\:636\:021$ & $270\:745$ & $1\:918\:833$ & $1\:882\:629$ \\ 
\bottomrule
\end{tabular}
\label{table:meshes_automatic_adaptation}
\end{table}
After executing each test separately (adapted tests) and aiming to investigate the effect of an oversampled mesh, we solved each frequency using a single mesh (i.e. the finest mesh which is that adapted to 2 Hz labeled H in Table \ref{table:meshes_automatic_adaptation}). Figures \ref{fig:total_field_X_model3} and \ref{fig:phase_field_X_model3} compare the amplitude $|\mathbf{E}_{x}|$ and phase $\Phi_{x}$ components of electric fields between PETGEM and those obtained by \citet{Weiss2006a}. Here, the overall agreement is better than those presented in Section \ref{DIPOLE1D_test}. However, simulation results with the oversampled mesh show higher differences with respect to the reference. Since these discrepancies have greater presence in the boundaries and around to source position, this accuracy loss is related to the quality of the elements on such mesh regions. 

\begin{figure}[!htbp]
	\centering
	\includegraphics[trim=0 55 0 0, clip, width=12cm]{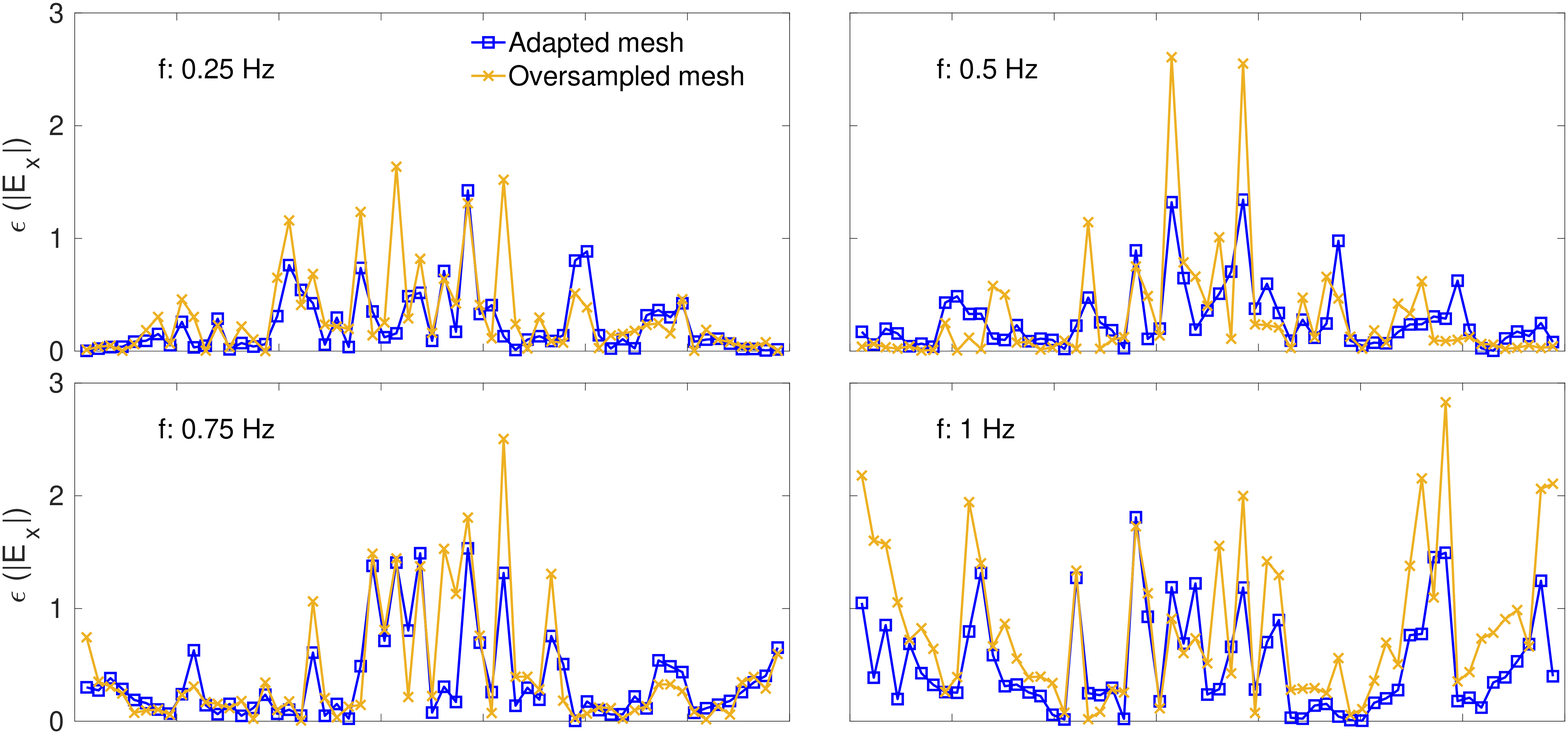}
	\includegraphics[width=12cm]{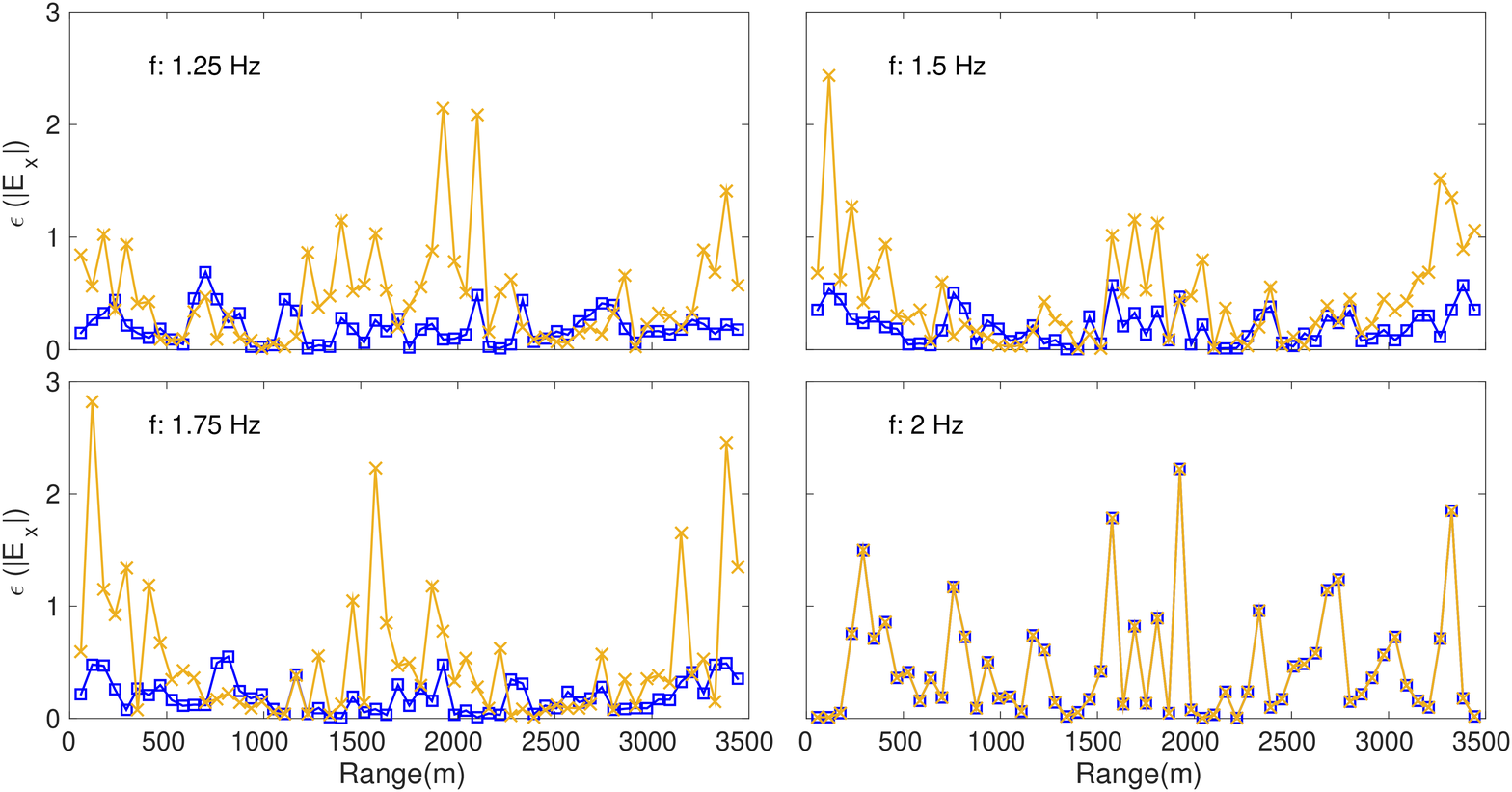}
	\caption{Amplitude misfits $\epsilon(|\mathbf{E}_{x}|)$ for Figure ~\ref{fig:total_field_X_model3}. An average relative misfit $<2\%$ is observed.}
	\label{fig:diff_amplitude_X_model3}
\end{figure}

\begin{figure}[!htbp]
	\centering
	\includegraphics[trim=0 55 0 0, clip, width=12cm]{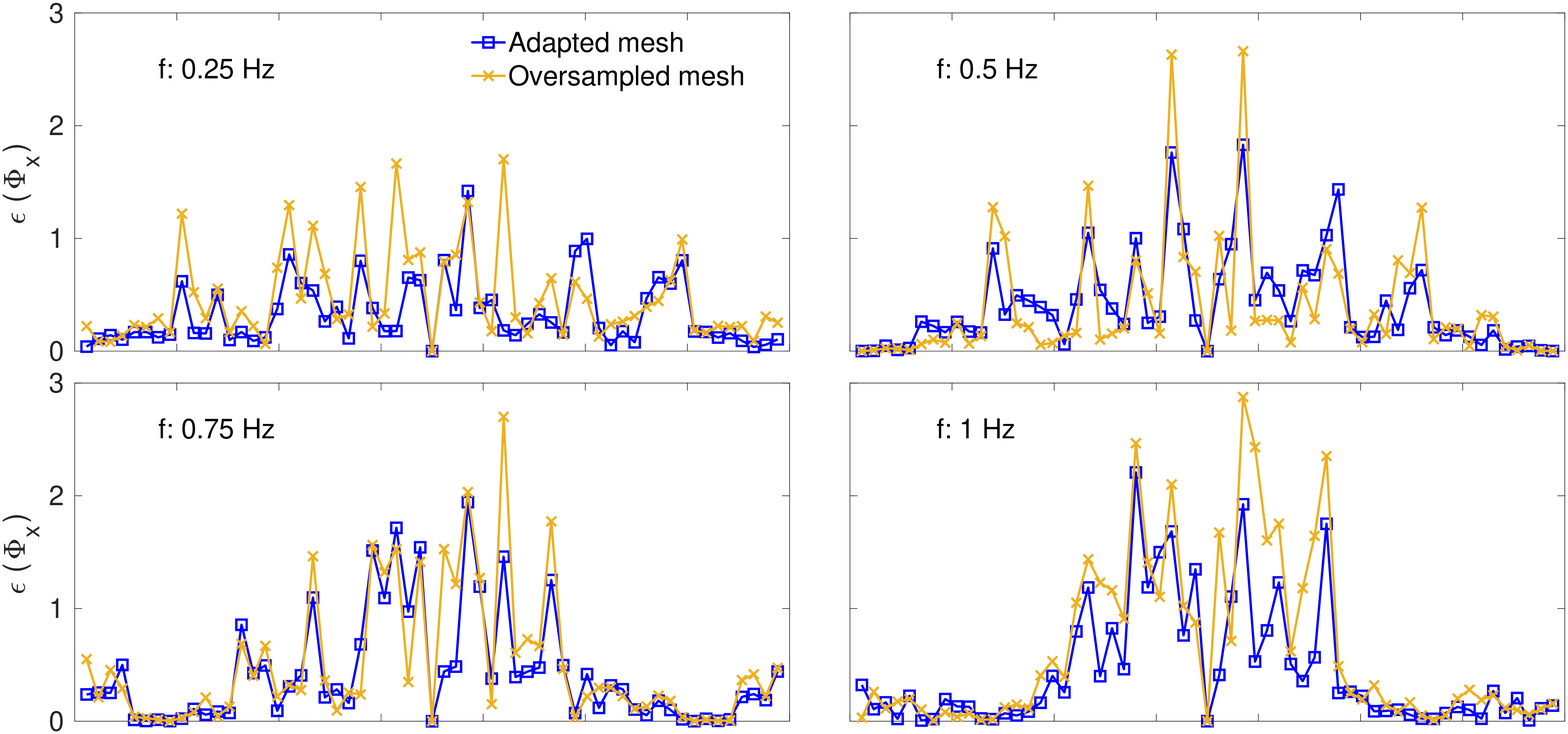}
	\includegraphics[width=12cm]{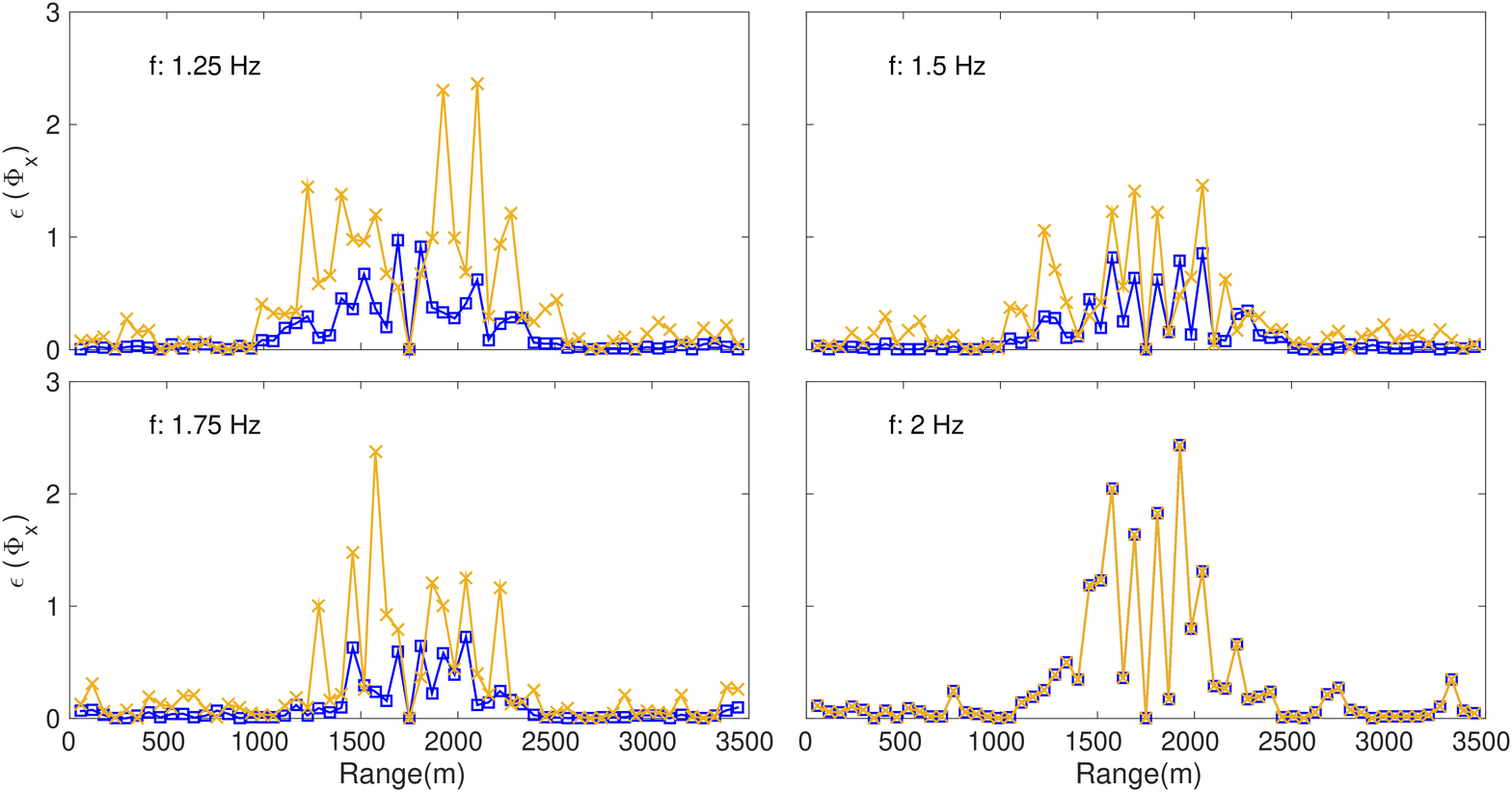}
	\caption{Phase misfits $\epsilon(\Phi_{x})$ of $\mathbf{E}_{x}$ solution in Figure ~\ref{fig:phase_field_X_model3}. Here, the $\epsilon(\Phi_{x})$ is plotted for each frequency, where an average relative misfit $<2\%$ is observed.}	
	\label{fig:diff_phase_X_model3}
\end{figure}

\begin{figure}[!htbp]
	\centering
	\includegraphics[width=12cm]{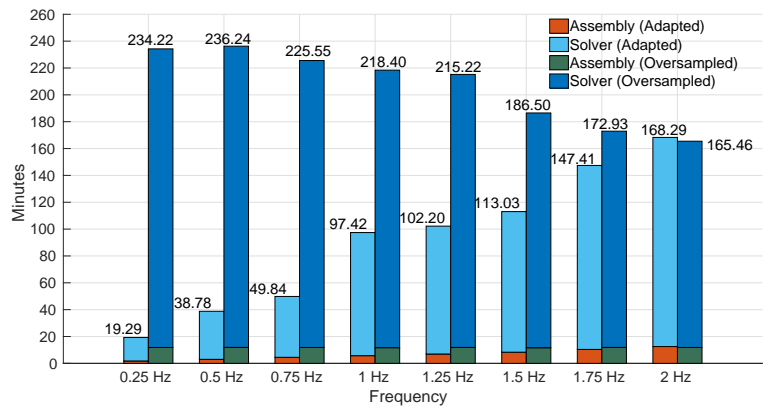}
	\caption{Time comparison between adapted and oversampled meshes. Assembly and solving times are considered.}
	\label{fig:timers_model3}
\end{figure}

We computed the amplitude ($\epsilon(|\mathbf{E}_x|)$) and phase ($\epsilon(\Phi_{x})$) misfits of the electric responses obtained on both meshes, adapted and oversampled. The $\epsilon(|\mathbf{E}_x|)$ and $\epsilon(\Phi_{x})$ misfits are shown in Figures \ref{fig:diff_amplitude_X_model3} and \ref{fig:diff_phase_X_model3}, respectively. One can clearly see the positive impact of frequency adapted meshes, e.g. the misfits associated to modeling results at 2 Hz are smaller than the ones shown in Figure \ref{fig:diff_X_model1}. The responses are very similar, the amplitude misfits are smaller than $2\%$, and the phase misfits are around $0.3$ rad. Notice that the results shown here, are a direct consequence of using unstructured tetrahedral meshes, which can be generated fully automatically given a proper geometry and spacing rules. This is different from conforming hexahedral grids that often require manual interaction to properly honor arbitrary spacings and geometries \citep{Owen1998}.
These numerical examples depict the usefulness of frequency-adapted meshing for 3D CSEM design and scenario studies. To illustrate the efficiency of this technique, we have measured the times for the assembly and solving tasks in PETGEM. Table \ref{table:summary_results_automatic_adaptation} lists the number of iterations, assembling and solution times on a single CPU. Solutions on oversampled meshes requires much more iterations than tests with adapted meshes and, as a consequence, need considerably more CPU time. On average, the adapted scheme is about four times faster in this set of experiments. In Figure \ref{fig:timers_model3}, assembly and solution times of both test sets are displayed. 
\begin{table}[!htbp]
\centering
\caption{Summary of results for mesh-adapted tests (M.A.) and oversampled mesh tests (O.S.). Times are expressed in minutes.}
\small
\begin{tabular}{ccccccc}
\toprule
\multirow{2}{2.6cm}{Frequency (Hz)} &  \multicolumn{2}{c}{Assembly} & \multicolumn{2}{c}{Solver} & \multicolumn{2}{c}{Iterations}\\
\cline{2-7} & {M.A.} & {O.S.}  & {M.A.} & {O.S.} & {M.A.} & {O.S.}\\
\midrule
0.25 & 1.77  & 11.8 & 17.52  & 222.4 & $5\:950$ & $11\:382$ \\
0.5  & 2.99  & 11.9 & 35.78  & 222.3 & $6\:630$ & $12\:519$ \\
0.75 & 4.49  & 11.8 & 45.35  & 213.7 & $5\:625$ & $9\:685$ \\
1    & 5.72  & 11.6 & 91.70  & 206.7 & $8\:245$ & $9\:275$ \\
1.25 & 6.95  & 11.8 & 95.70  & 203.4 & $6\:825$ & $8\:670$ \\
1.5  & 8.32  & 11.6 & 104.71 & 174.9 & $6\:145$ & $7\:540$ \\
1.75 & 10.38 & 11.8 & 137.02 & 161.1 & $7\:290$ & $7\:230$ \\
2    & 12.52 & 11.8 & 155.76 & 153.6 & $6\:935$ & $6\:935$ \\
\bottomrule
\end{tabular}
\label{table:summary_results_automatic_adaptation}
\end{table}
Previous results show the relevance of mesh adaptation for survey design in the 3D CSEM context.

\section{Scalability tests}
\label{Scalability tests}
Finally, we perform a series of tests to examine the scaling of PETGEM on distributed-memory architectures by running the same problem for different number of CPUs working in parallel. In this set of experiments the most time-consuming sections have been considered, namely, assembly and solving tasks. All simulations have been carried out on version III of the \textit{Marenostrum} (MN3) supercomputer at \textit{BSC}.

The following tests are based on the canonical model described in Section~\ref{DIPOLE1D_test}. Its mesh has been created with \textit{Gmsh}~\citep{Geuzaine2008} and has $35\:613\:880$ elements, $5\:947\:035$ nodes and $41\:753\:430$ edges. These tests have been carried out using 32, 64, 128, 256, 512, and $1\:024$ CPUs. These experiments are relevant because parallelism on distributed-memory platforms offers greater flexibility and capacity for realistic-scale 3D CSEM modeling.

\begin{figure}[!htbp]
\centering
\includegraphics[width=9cm]{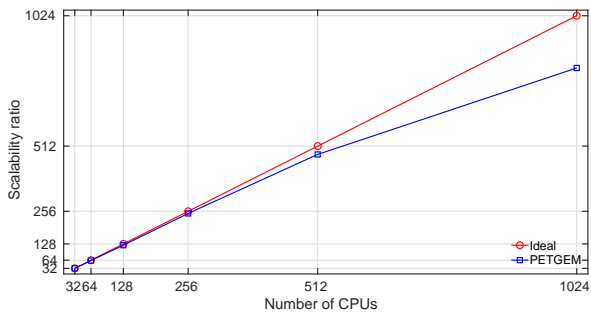}
\caption{Scalability tests for the second-level mesh ($41\:753\:430$ DOFs) on MN3.}
\label{fig:Scalability_distributed_memory_MNIII}
\end{figure}

Figure~\ref{fig:Scalability_distributed_memory_MNIII} shows speed-ups obtained for up to $1\:024$ CPUs of MN3 for case under consideration. The achieved scalability is almost linear for up to 256 CPUs. From this number on, the scalability stops its near-linear growth and slowly begins to saturate since the execution becomes dominated by exchange of messages between MPI tasks. However, the speed-ups keep growing constantly and significant reductions in runtime for more than thousand CPUs have been observed.
Table~\ref{table:summary_times_runtime_distributed_memory} shows the runtime, speed-up and parallel efficiency of the modeling. Analysing these results, it is easy to see that the computation time has been reduced by increasing the number of processes (around 26 times when increasing the number of CPUs from 32 to $1\:024$).
\begin{table}[htbp!]
\centering
\caption{Execution results for different number of CPUs on distributed-memory architectures}
\small
\begin{tabular}{ccccccc}
\toprule
$\#$ CPUs & 32 & 64 & 128 & 256 & 512 & $1\:024$\\ 
\midrule
Runtime (Min) & 945.30 & 482.14 & 246.09 & 122.96 & 63.08 & 36.92 \\
Speed-up & - & 1.96 & 3.84 & 7.68 & 14.98 & 25.60 \\
Efficiency & - & .98 & .96 & .96 & .94 & .80 \\
\bottomrule
\end{tabular}
\label{table:summary_times_runtime_distributed_memory}
\end{table}
In order to perform a more thorough analysis of the MPI parallelism within PETGEM, we have carried out a set of simulations that has been been analyzed using \textit{Paraver} \citep{Paraver2017}, which is a parallel trace analyzer developed at \textit{BSC}. This tool generates information that can be used in order to optimize a parallel application.

\begin{figure}[htbp!]
\centering
\includegraphics[width=9cm]{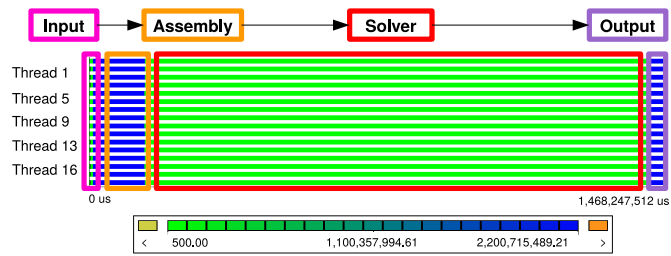}
\caption{Main computational phases in PETGEM. Here, the useful duration is plotted for each CPU as function of time.}
\label{fig:Code_structure_paraver}
\end{figure}

The analysis methodology begins with a set of \textit{Paraver} traces for a number of MPI tasks, obtained from executing an instrumented version of PETGEM. Then, from a visual analysis of the traces clean, cuts are generated in order to identify the main computational phases. This allows discerning main computational regions from communication intensive stages, e.g. elemental matrix computations or solving phase from MPI calls. Furthermore, in this step, additional information such as useful computational duration and number of MPI calls could be measured. The net result of this analysis phase is depicted in Figure~\ref{fig:Code_structure_paraver}, where the color represents the duration of the computation burst (useful duration). This view gives a good perception of where are the major computation phases, and their balance across processors. As a result, in Figure~\ref{fig:Code_structure_paraver} it is easy to see that assembly and solver phases are the main computational regions.
Once the code structure has been identified, we analyze \textit{Paraver} traces of PETGEM using 16, 32, and 64 CPUs in order to measure times for the aforementioned main computational regions. The main advantage of this strategy is that trace sizes are smaller and more manageable. In addition, all the effects that appear with the increase of the CPUs can be noticed much earlier.

\begin{figure}[htbp!]
\centering
\includegraphics[width=9cm]{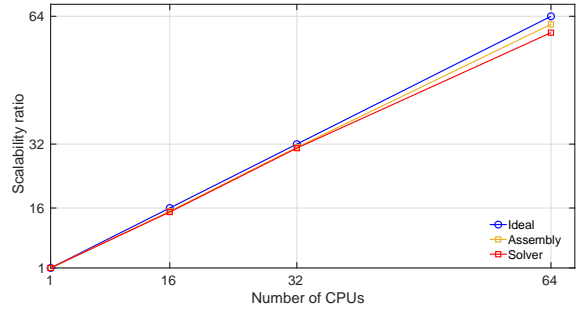}
\caption{Scalability ratio of main computational phases in PETGEM}
\label{fig:Scalability_distributed_memory_paraver_MNIII}
\end{figure}

\begin{figure}[htbp!]
\centering
\includegraphics[width=9cm]{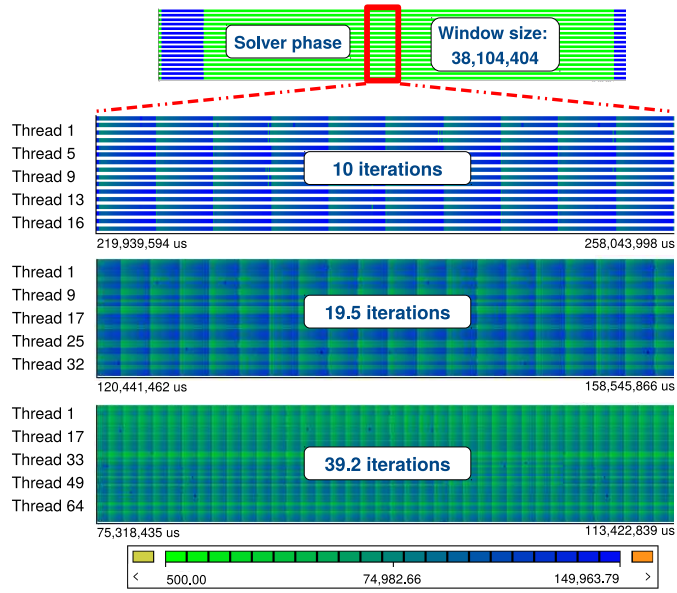}
\caption{Solver scalability analysis for different number of CPUs using Paraver.}
\label{fig:Paraver_traces}
\end{figure}

The scalability ratio for these experiments is shown in Figure~\ref{fig:Scalability_distributed_memory_paraver_MNIII}, where it is easy to observe a quasi-linear ratio for up to 64 CPUs. Furthermore, the assembly task is slightly more efficient than solving because it is an embarrassingly parallel task. For the sake of clarity, we analyze the number of solver iterations for each \textit{Paraver} trace. Thus, we cut a representative region of the solver phase and we measure the time (window size) for a number of iterations in the trace with 16 MPI tasks. Then, we count the number of iterations that fit in the same window size for remaining \textit{Paraver} traces (32 and 64 CPUs). In our experiments, we fixed the initial number of solver iterations to 10, which produced a window size equal to $38\:104\:404$ microseconds ($\mu s$). Results of this analysis are depicted in Figure~\ref{fig:Paraver_traces}, were it is easy to see an acceptable performance in terms of number of iterations when increasing number of CPUs. The examples show that PETGEM offers a good parallel performance for the solution of the problem under consideration.

\section{Conclusions}
\label{Conclusions}
We have presented PETGEM, a 3D CSEM modeling tool that has been designed to cope with the topic of computational geophysics. Its structure has been designed to cope simultaneously with three key requirements: accuracy, flexibility, and efficiency. The proposed code is based on the N\'ed\'elec Edge Finite Element Method (EFEM) and pure Python 3 language. This is to our knowledge the first time this type of approach has been applied for running geophysical simulations on HPC platforms. By design, the code is very modular and flexible, which could allow users to easily switch its underlying numerics or expand its capabilities. 

The accuracy and efficiency of this tool have been demonstrated on three different 3D CSEM models. The first model, a canonical test of an off-shore hydrocarbon reservoir, demonstrated a good agreement with the quasi-analytical solution. In the second model we have tested the code's capability to handle bathymetry. This model is especially interesting because it highlights one of the benefits of using an unstructured tetrahedral mesh, i.e. honoring complex geological structures without critically increasing the problem size. In the third test we exploit another advantage of unstructured tetrahedral meshes by building an automatic mesh adaptation technique for a multi-frequency study. This approach ensures, for a given frequency and source position, that the computational domain is consistent with the discretization of the EM equations with minimum amount of DOFs. For this simulation we observed that unnecessary and excessive refinement (over-sampled) examples required many more iterations than tests with adapted meshes, as a consequence needed considerably more CPU time. The adaptive mesh solutions had a factor of savings of up to four in time and storage compared to the fix-mesh result. Finally, a scalability study has shown that PETGEM offers an acceptable parallel performance for the physical problem under consideration.

Although PETGEM features will be beneficial for geophysicists interested in highly detailed and realistic 3D CSEM modeling,  we anticipate that this algorithm will prove more useful as a kernel for inversions of EM datasets. By combining a good code structure, simple programming language and relying on external and robust preprocessing and solving libraries, PETGEM  is a competitive, yet easy to use and expand, tool for 3D CSEM modeling. This effort, we hope, will foster research both in 3D CSEM modeling (and inversion) and in computer science.

\section{Code availability}
The PETGEM code is freely available at home page (\textit{http://petgem.bsc.es}), from PyPI repository (\textit{https://pypi.python.org/pypi/petgem/}), or by requesting the author (\textit{octavio.castillo@bsc.es}, \textit{ocastilloreyes@gmail.com}). In all cases, the code is supplied in a manner to ease the immediate execution under Linux platforms. User's manual and technical documentation (developer's guide) are provided in the PETGEM  archive as well.

\section{Acknowledgements}
The authors are acknowledge the support of the \textit{Mexican National Council for Science and Technology} (CONACyT). Octavio Castillo-Reyes expresses his gratitude to Prof. Yonghyun Chung of \textit{Seoul National University} for the model and synthetic data used in Figure \ref{fig:total_field_X_model2} and Figure \ref{fig:phase_field_X_model2}. Thanks to the \textit{BSC} (www.bsc.es) for allocating computation resources at \textit{Marenostrum} supercomputer. 

This project has received funding from the \textit{European Union\textquotesingle s Horizon 2020 research and innovation programme} under the \textit{Marie Sklodowska-Curie} grant agreement No. 644202. In addition, the research leading to these results has received funding from the \textit{European Union\textquotesingle s Horizon 2020 Programme} (2014-2020) and from \textit{Brazilian Ministry of Science, Technology and Innovation} through \textit{Rede Nacional de Pesquisa} (RNP) under the HPC4E Project (www.hpc4e.eu), grant agreement No. 689772.

The authors gratefully acknowledge (in alphabetically order) to Dr. Eduardo S\'anchez (\textit{BSC}) and Dr. Otilio Rojas (\textit{BSC}) for their valuable discussions and proofreading of this manuscript. Finally, last but not least, the authors are thankful to the reviewers for their valuable suggestions and comments.

\appendix
\section{Source term formulation}
\label{Appendix_A}
If we consider eq.~\eqref{eq:maxwell_diffusive_form1} and eq.~\eqref{eq:maxwell_diffusive_form2} with harmonic time dependence e$^{-i \omega t}$, $\omega$ as the angular frequency, $\mu_{0}$ as the free space magnetic permeability, $\mathbf{J}_{s}$ as the distribution of source current, for homogeneous media the source term for a x-directed dipole is given by
\begin{align}
\mathbf{E} &= \frac{I \cdot dS}{4 \pi \sigma r^{3}} \cdot e^{-ikr} \begin{bmatrix}
																(\frac{d_{\,x}^{2}}{r^{2}} \cdot -k^2 \cdot r^{2} + 3ikr + 3 k^{2} r^{2}-ikr-1) \cdot \hat{i} \\
																(\frac{d_{\,x} \cdot d_{\,y}}{r^{2}} \cdot -k^2 r^{2} + 3ikr + 3) \cdot \hat{j} \\
																(\frac{d_{\,x} \cdot d_{\,z}}{r^{2}} \cdot -k^2 r^{2} + 3ikr + 3) \cdot \hat{k}
														   \end{bmatrix}_{.}
\label{eq:x_directed_dipole}
\end{align}
For a y-directed dipole, the source term is defined by
\begin{align}
\mathbf{E} &= \frac{I \cdot dS}{4 \pi \sigma r^{3}} \cdot e^{-ikr} \begin{bmatrix}
																(\frac{d_{\,x} \cdot d_{\,y}}{r^{2}} \cdot -k^2 r^{2} + 3ikr + 3) \cdot \hat{i}\\
																(\frac{d_{\,y}^{2}}{r^{2}} \cdot -k^2 r^{2} + 3ikr + 3 + k^{2} r^{2}-ikr-1) \cdot \hat{j} \\
																(\frac{d_{\,y} \cdot d_{\,z}}{r^{2}} \cdot -k^2 r^{2} + 3ikr + 3) \cdot \hat{k} \\
														   \end{bmatrix}_{.}
\label{eq:y_directed_dipole}
\end{align}
Finally, following expression define the source term for a z-directed dipole
\begin{align}
\mathbf{E} &= \frac{I \cdot dS}{4 \pi \sigma r^{3}} \cdot e^{-ikr} \begin{bmatrix}
																(\frac{d_{\,x} \cdot d_{\,z}}{r^{2}} \cdot -k^2 \cdot r^{2} + 3ikr + 3) \cdot \hat{i}\\
																(\frac{d_{\,z} \cdot d_{\,y}}{r^{2}} \cdot -k^2 \cdot r^{2} + 3ikr + 3) \cdot \hat{j} \\
																(\frac{d_{\,z}^{2}}{r^{2}} \cdot -k^2 \cdot r^{2} + 3ikr + 3 + k^{2} r^{2}-ikr-1) \cdot \hat{k} \\
														   \end{bmatrix}_{.}
\label{eq:z_directed_dipole}
\end{align}
In eq.~\eqref{eq:x_directed_dipole}, eq.~\eqref{eq:y_directed_dipole} and eq.~ \eqref{eq:z_directed_dipole}, $I$ is the dipole current, $dS$ is the dipole length, $\sigma$ represent the background conductivity, $k$ is the propagation parameter (wavenumber), $r$ is the module between source position and evaluation point position, and $(d_{\,x},\: d_{\,y},\: d_{\,z})$ is the distance between source position and evaluation point position.

\section{N\'ed\'elec element basis} 
\label{Appendix_B}
If barycentric coordinates finite element defined in \citet{Zienkiewicz1977, Monk2003}  are renamed as $\boldsymbol{\lambda}^{e}_{i} = \mathbf{N}^{e}_{i}$, the vector basis functions for tetrahedral edge elements can be expressed in terms of first order of node-based Finite Element \citep{Jin2002}. Therefore, the vectorial functions for edge elements (see Figure \ref{fig:edge_element}) implemented in PETGEM  are defined by
\begin{align}
\mathbf{N}^{e}_{1} &= W_{12}\ell^{e}_{1} = (\boldsymbol{\lambda}^{e}_{1} \nabla \lambda^{e}_{2} - \boldsymbol{\lambda}^{e}_{2} \nabla \lambda^{e}_{1}) \ell^{e}_{1}, \label{eq:nedelec_basis_3D1}\\
\mathbf{N}^{e}_{2} &= W_{13}\ell^{e}_{2} = (\boldsymbol{\lambda}^{e}_{1} \nabla \lambda^{e}_{3} - \boldsymbol{\lambda}^{e}_{3} \nabla \lambda^{e}_{1}) \ell^{e}_{2}, \label{eq:nedelec_basis_3D2}\\
\mathbf{N}^{e}_{3} &= W_{14}\ell^{e}_{3} = (\boldsymbol{\lambda}^{e}_{1} \nabla \lambda^{e}_{4} - \boldsymbol{\lambda}^{e}_{4} \nabla \lambda^{e}_{1}) \ell^{e}_{3}, \label{eq:nedelec_basis_3D3}\\
\mathbf{N}^{e}_{4} &= W_{23}\ell^{e}_{4} = (\boldsymbol{\lambda}^{e}_{2} \nabla \lambda^{e}_{3} - \boldsymbol{\lambda}^{e}_{3} \nabla \lambda^{e}_{2}) \ell^{e}_{4}, \label{eq:nedelec_basis_3D4}\\
\mathbf{N}^{e}_{5} &= W_{42}\ell^{e}_{5} = (\boldsymbol{\lambda}^{e}_{4} \nabla \lambda^{e}_{2} - \boldsymbol{\lambda}^{e}_{2} \nabla \lambda^{e}_{4}) \ell^{e}_{5}, \label{eq:nedelec_basis_3D5}\\
\mathbf{N}^{e}_{6} &= W_{34}\ell^{e}_{6} = (\boldsymbol{\lambda}^{e}_{3} \nabla \lambda^{e}_{4} - \boldsymbol{\lambda}^{e}_{4} \nabla \lambda^{e}_{3}) \ell^{e}_{6}. \label{eq:nedelec_basis_3D6}
\end{align}
The gradients $\mathbf{\nabla} \lambda^{e}_{i}$ can be expanded as
\begin{align}
\nabla \lambda^{e}_{i} &= \dfrac{1}{6V^{e}} \begin{bmatrix}
b^{e}_{i} \cdot \hat{i} \\ 
c^{e}_{i} \cdot \hat{j} \\
d^{e}_{i} \cdot \hat{k} \end{bmatrix}_{.}
\end{align}
Therefore, the expanded form of the edge basis functions for tetrahedral edge elements in PETGEM  is the following
\begin{align}
\mathbf{W}_{12}\ell^{e}_{1} &= \dfrac{1}{(6V^{e})^{2}} \begin{bmatrix}
b^{e}_{2} ( a^{e}_{1} + b^{e}_{1}x + c^{e}_{1}y + d^{e}_{1}z ) - b^{e}_{1} ( a^{e}_{2} + b^{e}_{2}x + c^{e}_{2}y + d^{e}_{2}z ) \cdot \hat{i} \\
c^{e}_{2} ( a^{e}_{1} + b^{e}_{1}x + c^{e}_{1}y + d^{e}_{1}z ) - c^{e}_{1} ( a^{e}_{2} + b^{e}_{2}x + c^{e}_{2}y + d^{e}_{2}z ) \cdot \hat{j} \\
d^{e}_{2} ( a^{e}_{1} + b^{e}_{1}x + c^{e}_{1}y + d^{e}_{1}z ) - d^{e}_{1} ( a^{e}_{2} + b^{e}_{2}x + c^{e}_{2}y + d^{e}_{2}z ) \cdot \hat{k}
\end{bmatrix} \ell_{1}, \label{eq:nedelec_basis_3D7}\\
\mathbf{W}_{13}\ell^{e}_{2} &= \dfrac{1}{(6V^{e})^{2}} \begin{bmatrix}
b^{e}_{3} ( a^{e}_{1} + b^{e}_{1}x + c^{e}_{1}y + d^{e}_{1}z ) - b^{e}_{1} ( a^{e}_{3} + b^{e}_{3}x + c^{e}_{3}y + d^{e}_{3}z ) \cdot \hat{i} \\
c^{e}_{3} ( a^{e}_{1} + b^{e}_{1}x + c^{e}_{1}y + d^{e}_{1}z ) - c^{e}_{1} ( a^{e}_{3} + b^{e}_{3}x + c^{e}_{3}y + d^{e}_{3}z ) \cdot \hat{j} \\
d^{e}_{3} ( a^{e}_{1} + b^{e}_{1}x + c^{e}_{1}y + d^{e}_{1}z ) - d^{e}_{1} ( a^{e}_{3} + b^{e}_{3}x + c^{e}_{3}y + d^{e}_{3}z ) \cdot \hat{k}
\end{bmatrix} \ell_{2}, \label{eq:nedelec_basis_3D8}\\
\mathbf{W}_{14}\ell^{e}_{3} &= \dfrac{1}{(6V^{e})^{2}} \begin{bmatrix}
b^{e}_{4} ( a^{e}_{1} + b^{e}_{1}x + c^{e}_{1}y + d^{e}_{1}z ) - b^{e}_{1} ( a^{e}_{4} + b^{e}_{4}x + c^{e}_{4}y + d^{e}_{4}z ) \cdot \hat{i} \\
c^{e}_{4} ( a^{e}_{1} + b^{e}_{1}x + c^{e}_{1}y + d^{e}_{1}z ) - c^{e}_{1} ( a^{e}_{4} + b^{e}_{4}x + c^{e}_{4}y + d^{e}_{4}z ) \cdot \hat{j} \\
d^{e}_{4} ( a^{e}_{1} + b^{e}_{1}x + c^{e}_{1}y + d^{e}_{1}z ) - d^{e}_{1} ( a^{e}_{4} + b^{e}_{4}x + c^{e}_{4}y + d^{e}_{4}z ) \cdot \hat{k}
\end{bmatrix} \ell_{3}, \label{eq:nedelec_basis_3D9}\\
\mathbf{W}_{23}\ell^{e}_{4} &= \dfrac{1}{(6V^{e})^{2}} \begin{bmatrix}
b^{e}_{3} ( a^{e}_{2} + b^{e}_{2}x + c^{e}_{2}y + d^{e}_{2}z ) - b^{e}_{2} ( a^{e}_{3} + b^{e}_{3}x + c^{e}_{3}y + d^{e}_{3}z ) \cdot \hat{i} \\
c^{e}_{3} ( a^{e}_{2} + b^{e}_{2}x + c^{e}_{2}y + d^{e}_{2}z ) - c^{e}_{2} ( a^{e}_{3} + b^{e}_{3}x + c^{e}_{3}y + d^{e}_{3}z ) \cdot \hat{j} \\
d^{e}_{3} ( a^{e}_{2} + b^{e}_{2}x + c^{e}_{2}y + d^{e}_{2}z ) - d^{e}_{2} ( a^{e}_{3} + b^{e}_{3}x + c^{e}_{3}y + d^{e}_{3}z ) \cdot \hat{k}
\end{bmatrix} \ell_{4}, \label{eq:nedelec_basis_3D10}\\
\mathbf{W}_{42}\ell^{e}_{5} &= \dfrac{1}{(6V^{e})^{2}} \begin{bmatrix}
b^{e}_{2} ( a^{e}_{4} + b^{e}_{4}x + c^{e}_{4}y + d^{e}_{4}z ) - b^{e}_{4} ( a^{e}_{2} + b^{e}_{2}x + c^{e}_{2}y + d^{e}_{2}z ) \cdot \hat{i} \\
c^{e}_{2} ( a^{e}_{4} + b^{e}_{4}x + c^{e}_{4}y + d^{e}_{4}z ) - c^{e}_{4} ( a^{e}_{2} + b^{e}_{2}x + c^{e}_{2}y + d^{e}_{2}z ) \cdot \hat{j} \\
d^{e}_{2} ( a^{e}_{4} + b^{e}_{4}x + c^{e}_{4}y + d^{e}_{4}z ) - d^{e}_{4} ( a^{e}_{2} + b^{e}_{2}x + c^{e}_{2}y + d^{e}_{2}z ) \cdot \hat{k}
\end{bmatrix} \ell_{5}, \label{eq:nedelec_basis_3D11}\\
\mathbf{W}_{34}\ell^{e}_{6} &= \dfrac{1}{(6V^{e})^{2}} \begin{bmatrix}
b^{e}_{4} ( a^{e}_{3} + b^{e}_{3}x + c^{e}_{3}y + d^{e}_{3}z ) - b^{e}_{3} ( a^{e}_{4} + b^{e}_{4}x + c^{e}_{4}y + d^{e}_{4}z ) \cdot \hat{i} \\
c^{e}_{4} ( a^{e}_{3} + b^{e}_{3}x + c^{e}_{3}y + d^{e}_{3}z ) - c^{e}_{3} ( a^{e}_{4} + b^{e}_{4}x + c^{e}_{4}y + d^{e}_{4}z ) \cdot \hat{j} \\
d^{e}_{4} ( a^{e}_{3} + b^{e}_{3}x + c^{e}_{3}y + d^{e}_{3}z ) - d^{e}_{3} ( a^{e}_{4} + b^{e}_{4}x + c^{e}_{4}y + d^{e}_{4}z ) \cdot \hat{k}
\end{bmatrix} \ell_{6}. \label{eq:nedelec_basis_3D12}
\end{align}
Eqs.~\eqref{eq:nedelec_basis_3D7} - \eqref{eq:nedelec_basis_3D12} are divergence free but not curl free \citep{Jin2002}.

We redefine the stiffness matrix and mass matrix by \citet{Jin2002} as follows
\begin{align}
K^{e}_{ij} &= \iiint_{V^{e}} (\boldsymbol{\nabla} \times \mathbf{N}^{e}_{i} \cdot S^{e}_{i})  \cdot (\boldsymbol{\nabla} \times \mathbf{N}^{e}_{j} \cdot S^{e}_{j}) \mspace{3mu} d V, \label{eq:stiffness_matrix} \\
M^{e}_{ij} &= \iiint_{V^{e}} (\mathbf{N}^{e}_{i} \cdot S^{e}_{i}) \cdot (\mathbf{N}^{e}_{j} \cdot S^{e}_{j})\mspace{3mu} d V, \label{eq:mass_matrix}
\end{align}
where, $S^{e}_{i}$ are coefficients equal $1$ or $-1$ depending on the local and global direction of the \textit{i-th} edge in the element \textit{e}. In PETGEM  these coefficients are computed as follows. If an edge adjoins two nodes $n_{i}$ and $n_{j}$, the direction of the edge is going from node $n_{i}$ to node $n_{j}$ if $i < j$. This simple algorithm gives a unique orientation of each edge in the mesh. On the other hand, the local orientation of edges within each element can be determined by his nodes indexes. Therefore, $S^{e}_{i}$ are given by the following vectorial function
\begin{align}
S^{e}_{i} = \frac{node^{e}_{i2} - node^{e}_{i1}}{|node^{e}_{i2} - node^{e}_{i1}|} & \quad \quad i=1\ldots 6,
\label{eq:signs_edges}
\end{align}
where $i$ is the edge index within \textit{e}-th element that adjoins $node^{e}_{i1}$ with $node^{e}_{i2}$. The main advantage of eq.~\eqref{eq:signs_edges} is that it allows to work with node numbering based on a clockwise or counter-clockwise in order to meet some conditions of FEM formulations such as element's volume computation, which must be positive in any case.

\section*{References}
\bibliography{references}

\end{document}